\documentclass[useAMS,usenatbib]{mn2e}

\usepackage{epsfig}
\usepackage{graphicx}
\usepackage[latin1]{inputenc}
\def\simlt{\ \raise -2.truept\hbox{\rlap{\hbox{$\sim$}}\raise5.truept   %
\hbox{$<$}\ }}
\def\simgt{\ \raise -2.truept\hbox{\rlap{\hbox{$\sim$}}\raise5.truept   %
\hbox{$>$}\ }}                                                          %

\def\be{\begin{equation}}
\def\ee{\end{equation}}
\def\newline{\hfil\break}

\def\la{\mathrel{\hbox{\rlap{\hbox{\lower4pt\hbox{$\sim$}}}\hbox{$<$}}}}
\def\ga{\mathrel{\hbox{\rlap{\hbox{\lower4pt\hbox{$\sim$}}}\hbox{$>$}}}}

\title[Radio emission in the Bullet cluster]{Is the radio emission in the Bullet cluster due to Dark Matter annihilation?}

\author[P. Marchegiani and S. Colafrancesco]
{P. Marchegiani$^{1}$\thanks{E-mail: Paolo.Marchegiani@wits.ac.za} and S. Colafrancesco$^{1}$\thanks{E-mail: Sergio.Colafrancesco@wits.ac.za}\\
$^{1}$School of Physics, University of the Witwatersrand, Private Bag 3, 2050-Johannesburg, South Africa
}
\begin{document}

\date{Accepted 2015 June 17. Received 2015 June 15; in original form 2015 April 20.}

\pagerange{\pageref{firstpage}--\pageref{lastpage}} \pubyear{2015}

\maketitle

\label{firstpage}

\begin{abstract}
We study the complex structure of the Bullet cluster radio halo to determine the Dark Matter (DM) contribution to the emission observed in the different subhalos corresponding to the DM and baryonic dominated regions. We use different non-thermal models to study the different regions, and we compare our results with the available observations in the radio, X-ray and gamma-ray bands, and the Sunyaev-Zel'dovich (SZ) effect data. We find that the radio emission coming from the main DM subhalo can be produced by secondary electrons produced by DM annihilations. In this scenario there are however some open issues, like the difficulty to explain the observed flux at 8.8 GHz, the high value of the required annihilation cross section, and the lack of observed emission coming from the minor DM subhalo. We also find that part of the radio emission originated by DM annihilation could be associated with a slightly extended radio source present near the main DM subhalo. Regarding the baryonic subhalos, the radio measurements do not allow to discriminate between a primary or secondary origin of the electrons, while the SZ effect data point towards a primary origin for the non-thermal electrons in the Main Subcluster. We conclude that in order to better constrain the properties of the DM subhalos, it is important to perform detailed measurements of the radio emission in the regions where the DM halos have their peaks, and that the separation of the complex radio halo in different subhalos is a promising technique to understand the properties of each specific subhalo.
\end{abstract}


\section{Introduction}

Several observations suggest that most of the matter in the universe is dark and non-baryonic (see, e.g., Bertone et al. 2005 for a review, and references therein). Various methods based on direct or indirect detection have been proposed to constrain the nature and the properties of the Dark Matter (DM). A promising way to study the DM properties consists in the detection of the electromagnetic emissions coming from astrophysical objects as a consequence of DM particles annihilation and secondary particle $e^{\pm}$ production, because the spectral features of these emissions are expected to be closely correlated with the nature, the composition and the mass of the DM particles (see, e.g., Colafrancesco et al. 2006). 

A recent study (Colafrancesco et al. 2015) showed that many cosmic structures with different size and mass can be promising candidates to detect the DM signal, especially in the radio band, using future high-sensitivity experiments like the Square Kilometre Array (SKA)  (e.g., Dewdney et al. 2012). Among these structures, dwarf spheroidal galaxies appear to be good candidates because in these objects the DM component is supposed to be dominant with respect to other diffuse components of baryonic origin (e.g., Regis et al. 2014), whereas in galaxy clusters the DM signal is expected to be stronger, but other mechanisms of baryonic origin can produce emissions superposed to the DM emission (see, e.g., Colafrancesco et al. 2010 for the case of the Perseus cluster).

Non-thermal emission in the atmospheres of galaxy clusters manifests itself with a wide range of emission mechanisms and can be then observed over a wide range of frequencies.
Synchrotron emission from electrons either primarily accelerated at shocks (Tribble 1993) or re-accelerated by turbulence (Brunetti et al. 2001), or secondarily produced by proton-proton collisions (Blasi \& Colafrancesco 1999) or DM annihilation (Colafrancesco \& Mele 2001, Colafrancesco et al. 2006, 2011b) is expected to be observed in the radio band.
The same primary and/or secondary electrons are also expected to emit by inverse Compton Scattering (ICS) off the Cosmic Microwave Background (CMB) photons (and possibly other cosmic backgrounds), and by non-thermal bremsstrahlung off the nuclei of the thermal gas over a wide energy range from ultraviolet to gamma-rays (e.g. Blasi \& Colafrancesco 1999).
A strong gamma ray emission, due to the $\pi^0 \to \gamma \gamma$ decay, is expected in secondary models, either produced by $p-p$ collisions or by DM annihilation (e.g., Colafrancesco \& Blasi 1998, Colafrancesco et al. 2006).
Other astrophysical effects related to these non-thermal components are also expected, like a non-thermal Sunyaev-Zel'dovich effect (SZE) (e.g. Colafrancesco et al. 2003) and gas heating by Coulomb and hadronic interactions (e.g. Colafrancesco \& Marchegiani 2008).

For these reasons, the detection of genuine DM-produced signal in galaxy clusters appears to be not easy, especially when the baryonic and the DM emissions are co-spatially located. Therefore, a good way to search for the DM emission is to study galaxy clusters where the DM and the baryonic halos are spatially offset. The most famous cluster with these properties is the Bullet cluster (1E 0657-558), where observations of the X-ray emission (e.g. Markevitch et al. 2002) and of the gravitational lensing (Clowe et al. 2006, Bradac et al. 2006) showed that distributions of the hot gas in the Intra Cluster Medium (ICM) and of the DM are centered at different positions.
The other case of A520 where there is some spatial offset between DM and baryons is not a very clean case to study (Mahdavi et al. 2007, Clowe et al. 2012).

The Bullet cluster is a merging cluster which shows a complex structure and, because of the spatial offset between hot baryons, DM and galaxy distributions, appears to be one of the best laboratories to decipher the stratification of the complex electron population.
In the X-rays this cluster shows the presence of two subclusters: a Main Subcluster (MS), with an elliptical shape elongated in the North-South direction, and a Bullet Subcluster (BS), located on the west of the MS, with a more asymmetric shape that shows a front shock in the west direction. This is currently interpreted by assuming that the cluster is in a post-merging phase, and the BS has just crossed the MS, producing gas heating and particles acceleration. The gravitational lensing analysis of the Bullet cluster shows two main halos: the larger one is located on the east of the MS, and the smaller one is located on the west of the BS. The offset between the hot gas and DM halos is explained assuming that this is a consequence of the recent merging where the collisional (the baryons) and the collisionless (DM) components have been separated after the powerful merging event. Finally, the spatial distribution of the galaxies in the cluster seems to be related more with the DM distribution rather than with the hot gas distribution.

Radio observations of the Bullet cluster (Liang et al. 2000, Shimwell et al. 2014) show a very complex morphology, with an elliptical radio halo elongated in the east-west direction, a large radio relic on the east of the halo (Shimwell et al. 2015), and several mini halos around galaxies and point-like sources. Interestingly, the overall radio halo, after the sources removal procedure (see Shimwell et al. 2014), appears to have three different peaks, two of which are located near the X-rays peaks, and another one located near the DM Eastern (DME) region. 
The spectrum of the radio halo also shows very interesting features: while the emission from the internal part of the halo, where the MS is located, has an almost perfect power-law shape, the spectrum of the extended region shows a flattening of the spectrum at $\nu\simgt 2.3$ GHz, and a steepening at $\nu\simgt 5$ GHz. This fact can be produced by the presence of different emission components  with different spectra that dominate alternately in different frequency ranges, one of which can be due to the DM component. 

For this reason, it is interesting to study in detail the emissions coming from the different regions of the radio halo and explore the possibility to separate the baryonic and the DM-produced emissions. This goal requires to proceed with a very detailed study of the different halos, using appropriate models for each of them, starting from radio emission and using, when available, information obtained in other spectral bands.\\
In fact, due to the degeneracy between electrons density and magnetic field intensity arising from radio observations (e.g., Longair 1994, Colafrancesco et al. 2005), it is important to observe, or constrain, the non-thermal emission in the Bullet cluster in other spectral bands, like the X-ray and gamma-ray bands. Unfortunately, only upper limits have been obtained so far. The most recent NuSTAR observations in the hard X-rays band obtained with an integration time of 266 ks (Wik et al. 2014) provided an upper limit $F(50-100 \mbox{ keV})<1.1\times10^{-12}$ erg cm$^{-2}$ s$^{-1}$. In the gamma ray energy range, an upper limit has been derived by the Fermi-LAT experiment, $F(0.2-100 \mbox{ GeV})<2.75\times10^{-9}$ cm$^{-2}$ s$^{-1}$ (Ackermann et al. 2010). 
The same authors provided in their Figure 1 a gamma ray upper limit, extrapolated to $E > 0.1$ GeV, that is $F(>0.1 \mbox{ GeV})<5.7\times10^{-9}$ cm$^{-2}$ s$^{-1}$.

Another source of information on the properties of the Bullet Cluster is provided by the SZE observations; the study of its spectral shape, especially at high frequencies ($\nu>300$ GHz), where the relativistic corrections are more relevant (Colafrancesco et al. 2003, Colafrancesco et al. 2011a), allows to obtain information on the thermal and possibly the non-thermal electrons in the cluster atmosphere. In the Bullet cluster, the data at high frequencies obtained with Herschel-SPIRE (Zemcov et al. 2010) allowed to establish that, in addition to main thermal component, there is a second electron population (more likely of non-thermal origin) that is contributing to the total observed SZE (Colafrancesco et al. 2011a), and that the gas temperature distribution along the line of sight is strongly inhomogeneous (Prokhorov \& Colafrancesco 2012).

By using all this information, we perform in this paper a detailed study of the Bullet cluster, starting from the spectral and the spatial properties of the radio halo, and using different emission models that are appropriate for the different regions. We also check the information obtained from the radio band considering the corresponding emissions in the high-energy (i.e., X-ray and gamma ray) bands, comparing the predictions with the available upper limits and with the sensitivities of coming and planned instruments. We also study the SZE in a more detailed way than in previous studies, and make predictions for future observations with instruments like Millimetron, a planned cooled 10 m space telescope which will operate in the Lagrangian L2 point in the 20 $\mu$m -- 20 mm wavelength range (Smirnov et al. 2012; Kardashev et al. 2015).

The outline of the paper is the following: in Sect. 2 we study the structure of the radio halo emission in the Bullet cluster with specific interest to the various sub-peaks of this emission. We discuss in Sect.3 the consequences of the radio halo analysis on the DM and CR distribution models, the possible issues, and the predictions in the high-energy bands. We then focus in Sect.4 on the analysis of the SZE. We finally summarize our results and draw our conclusions in Sect.5.\\
Throughout the paper, we use a flat, vacuum--dominated cosmological model following the results of Planck, with $\Omega_m = 0.308$, $\Omega_{\Lambda} = 0.692$ and $H_0 =67.8$ km s$^{-1}$ Mpc$^{-1}$ (Ade et al. 2015). With these values the luminosity distance for the Bullet cluster at $z=0.296$ is  $D_L=1576$ Mpc, and 1 arcmin corresponds to 273 kpc at this distance.

\section{The structure of the radio halo}

The observed radio halo of the Bullet cluster has an elliptical shape with the major axis oriented in the East-West direction (see, e.g., Fig.5 in Shimwell et al. 2014). Actually, the radio brightness inner contours of the radio halo suggest that its overall elliptical shape can be due to the combination of emissions of different subhalos; in particular, the main peak is located near the center of the MS X-ray emission, and the radio emission appears to be elongated both in the west direction, with a secondary peak located near the BS X-ray peak, and in the south-east direction, with another peak located near the peak of the DME distribution. A little residual emission seems to be present also near some of the radio sources identified in the Bullet cluster, as the sources indicated with A, G, H, F and K in Shimwell et al. (2014).

It is therefore justified to test the hypothesis that the overall radio emission is mainly due to the combination of three distinct radio sub-halos. In this framework we will first test some hypotheses on the origin of the radio halo emission that are consistent with both the available observations and the theoretical expectations, and then we will verify the consequences of these hypotheses based on their consistency with the available observations.

The hypothesis on the origin of the Bullet cluster radio halo that we want to test is the following:\\
\textit{i)} the sub-halos coincident with the MS and the BS are of baryonic origin; we assume that they are produced by a secondary electron model (SEM) of hadronic origin where the CR protons are normalized according to the Warming Ray (WR) model (see Colafrancesco \& Marchegiani 2008 for details) that is able to reproduce the X-ray temperature distribution of the two baryonic halos. The use of SEM seems to be reasonable because in this model the strongest emission comes from the regions where the ICM density is highest; the normalization of the non-thermal protons following the WR model provides actually an upper limit on their density, because a larger density would provide a stronger gas heating and therefore a higher temperature than  the observed one;\\
\textit{ii)} the radio sub-halos coincident with the DM halos are produced by secondary electrons produced by DM annihilation; we will focus on the case where the DM is constituted by neutralinos, i.e. the lightest particles in the minimal super-symmetric extension of the standard model (MSSM).

\subsection{The Main Subcluster}

The main subcluster (MS) is found at the location where the X-ray emission has its main peak (Markevitch et al. 2002), and at the same location also the radio brightness maps show a peak coincident with the X-ray one (Shimwell et al. 2014). 

The thermal gas temperature in this region is $kT=14.2^{+0.3}_{-0.2}$ keV (Wik et al. 2014), and the gas density profile can be approximated by an isothermal $\beta$-model:
\begin{equation}
n_{th}(r)=n_0 \left[1+\left(\frac{r}{r_c}\right)^2\right]^{-3\beta/2}
\end{equation}
with $n_0=7.1\times10^{-3}$ cm$^{-3}$, $\beta=1.04$ and $r_c=516$ kpc (Ota \& Mitsuda 2004), where the approximation of spherical symmetry is used.
The spatial extent of the radio emission in this region corresponds to the smallest of the regions considered in the analysis of Liang et al. (2000) (region \# 3 in their Figure 4), and the relative radio spectrum is the lower flux spectrum plotted in their Figure 7, which shows a radio spectral index of  $\alpha_R\sim1.4$.

When we apply the WR model to this subcluster, in which we assume that the spatial profile of non-thermal protons is $\propto n_{th}(r)^\alpha$, we find that this model requires  $\alpha=1$, consistently with many other isothermal clusters (see the analysis of Colafrancesco \& Marchegiani 2008 for details). 
Assuming a proton spectrum of the type
\begin{equation}
N_p(\gamma)=N_{p,0}\gamma^{-s_p}
\end{equation}
with $s_p=2.7$, the WR model provides the value $N_{p,0}=4.1\times10^{-8}$ cm$^{-3}$ for the CR proton normalization, which implies a ratio between the CRs and the thermal gas pressures at the cluster center of $P_{CR}/P_{th}=0.42$.

\begin{figure}
\begin{center}
{
 \epsfig{file=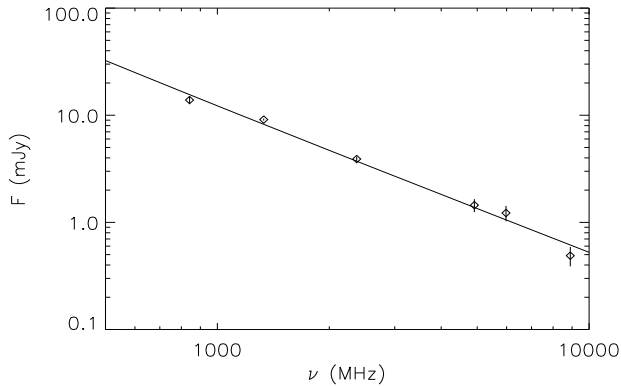,height=5.7cm,width=8.5cm,angle=0.0}
}
\end{center}
 \caption{The radio emission in the region of the Main Subcluster fitted with a WR model with $s_p=2.7$ and $B=7.5$ $\mu$G. The data shown in this plot correspond to the central region of the cluster as given by Liang et al. (2000).}
 \label{radio_central}
\end{figure}

The radius of the X-ray emission in this region is $R\sim520$ kpc, but, to determine the value of the magnetic field necessary to produce the observed radio emission, we calculate the emission produced inside a radius of $\sim160$ kpc, from the size of region \# 3 of Liang et al. (2000). 
In this way, if we assume that the magnetic field is uniform throughout this region, we find that the diffuse radio emission is reproduced by secondary electrons emitting in a magnetic field with amplitude  $B=7.5$ $\mu$G. 
Such value is compatible with the overall findings in other clusters (e.g., Carilli \& Taylor 2002), and it is reasonable in a strong merging cluster like the Bullet cluster which is very hot, with a highly non-uniform gas distribution where the recent shock activity could have amplified the pre-existing magnetic field (see also Shimwell et al. 2015). Figure \ref{radio_central} shows that this model can indeed reproduce quite well the data observed in the central region studied by Liang et al. (2000).

The hard X-ray flux produced in the 50--100 keV band by the secondary electrons in the WR model is $F(50-100 \mbox{ keV})=1.8\times10^{-16}$ erg cm$^{-2}$ s$^{-1}$, much lower than the NuSTAR limit (Wik et al. 2014). 
The gamma ray flux at $E>0.1$ GeV, mainly produced by the $\pi^0 \to \gamma \gamma$ channel, expected in this model is  $F(E>0.1 \mbox{ GeV}) = 5.1\times10^{-10}$ cm$^{-2}$ s$^{-1}$, well below the Fermi-LAT limit reported by Akermann et al. (2010).

This model for the MS is therefore fully consistent with all multi-frequency data available for the cluster, from radio to gamma-rays.

\subsection{The Bullet Subcluster}

The BS is centered on the region of the second peak in the X-ray emission, i.e. the one  that corresponds to the ``bullet" that crossed the MS and is producing the shock front.

This subcluster is colder in its central part (with a temperature of $kT\sim$ 7 keV), and it is hotter at its periphery 
(where the temperature reaches $kT\sim18$ keV), according to the analysis of Markevitch et al. (2002). The same authors found that the gas density is $n_{th} \sim1.7\times10^{-2}$ cm$^{-3}$ in the cold region, within a radius of $\sim50$ kpc from the center of the BS, and is $n_{th} \sim4.1\times10^{-3}$ cm$^{-3}$ in the hotter region, within a radius of $\sim260$ kpc from the BS center; in both regions it is found that the density is approximately constant with radius.  We can therefore consider that the BS has a structure similar to a cool-core cluster. 

When we apply the WR model to the BS we obtain a value $\alpha=0.39$. For a value of the protons spectral index $s_p=2.7$, the normalization of the proton spectrum required by the model is $N_{p,0}=4.9\times10^{-8}$ cm$^{-3}$ ($P_{CR}/P_{th}=0.46$), while for $s_p=2.9$ we find $N_{p,0}=5.8\times10^{-8}$ cm$^{-3}$ ($P_{CR}/P_{th}=0.67$).

We use this model to describe the radio halo data in the region \# 2 in Figure 4 of Liang et al. (2000), which is similar in extension to the smaller region considered in the analysis of Shimwell et al. (2014), and hence we use the higher level radio flux data in the spectrum obtained by  Liang et al. (2000), i.e. those in their Figure 7, and the lower amplitude radio flux spectrum reported in Figure 7 of Shimwell et al. (2014). 
The region \# 2 in Liang et al. (2000) is actually larger than the BS, and includes also the DME region and some radio sources. So, now we compare the radio emission produced in the BS with the region \# 2 data, but in the following we will consider the contributions from other sources in this region.
The data sets of Liang et al. (2000) and those of Shimwell et al. (2014) show that the radio halo spectrum in this region has a power-law shape with index  $\alpha_R\sim1.6$ up to a frequency of  $\nu\sim2.3$ GHz, while at higher frequencies the spectrum flattens between 2.3 and 5 GHz with a further steepening at frequencies $\nu \ge 5$ GHz. This spectral shape might suggest that the observed spectrum is due to the superposition of different components, one of which can be the BS, and a second one is flatter and dominates the overall spectrum at $2.3 \simlt \nu \simlt 5$ GHz, and then steepens at higher frequencies. 

\begin{figure}
\begin{center}
{
 \epsfig{file=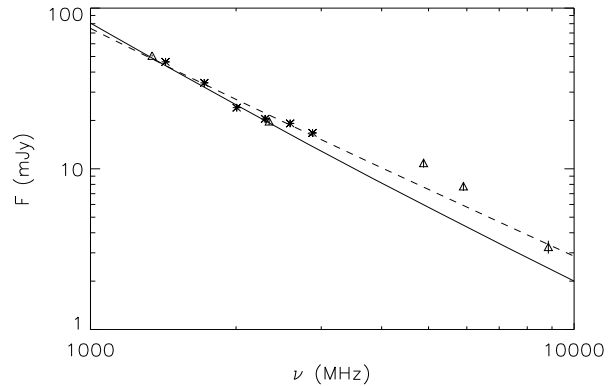,height=5.7cm,width=8.5cm,angle=0.0}
}
\end{center}
 \caption{The radio emission in the region of the Bullet Subcluster fitted with two WR models with $s_p=2.9$ and $B=60$ $\mu$G (solid line) and with $s_p=2.7$ and $B=27$ $\mu$G (dashed line).
 The data shown in this plot correspond to the region \# 2 of the cluster as given by Liang et al. (2000) (triangles) and to the smaller region considered by Shimwell et al. (2014) (asterisks).}
 \label{radio_bs}
\end{figure}

In Fig.\ref{radio_bs} we show the radio spectrum produced in the BS for the two considered WR models, with a magnetic field of $B=60$ $\mu$G for $s_p=2.9$, and with $B=27$ $\mu$G for $s_p=2.7$.
These values of the magnetic field are clearly extreme, so probably the BS emission alone is not sufficient to produce the flux observed in the whole region \# 2 of Liang et al. (2000); anyway, the BS is located near the shock region, and it hosts a cool core, hence a relatively high value of the magnetic field is conceivable. 
As we can see, the first model reproduces well the data up to a frequency of 2.3 GHz, and fails to reproduce the data at higher frequencies, while the second model is less accurate at small frequencies but, given the large data dispersion, it can explain the observed spectrum at all the frequencies.

The ICS flux produced in the 50--100 keV band is $4.1\times10^{-19}$ erg cm$^{-2}$ s$^{-1}$ for $s_p=2.9$, and $3.9\times10^{-18}$ erg cm$^{-2}$ s$^{-1}$ for $s_p=2.7$, both much lower than the limit of Wik et al. (2014).

The gamma ray flux for $E>0.1$ GeV produced by $\pi^0$ decay for $s_p=2.7$ is $9.7\times10^{-11}$ cm$^{-2}$ s$^{-1}$, and for $s_p=2.9$ is  $9.6\times10^{-11}$ cm$^{-2}$ s$^{-1}$, both below the limit of Ackermann et al. (2010).

\subsection{The Dark Matter Eastern peak}

Gravitational lensing image of the gravitational potential of the Bullet cluster shows two peaks separated from the X-ray peaks (Clowe et al. 2006), that are supposed to be dominated by the Dark Matter density distribution.
The luminous component (i.e., the galaxies) seems to follow the DM distribution rather than the thermal gas distribution.
The DME peak is found to be the most massive one, with a mass $M_{DME}\sim5.6\times10^{14} M_\odot$ within a region of 520 kpc radius (see Bradac et al. 2006).

Following the procedure described in Colafrancesco et al. (2015), we study the case where the DM is constituted by neutralinos with different values of the mass and compositions, and we use the relations between the halo mass and the parameters of the DM distribution, as the concentration parameter and the central density, and, for a Navarro, Frenk \& White (1996) mass density distribution,
\begin{equation}
\rho(r)=\frac{\rho_s}{\left(\frac{r}{r_s}\right)\left(1+\frac{r}{r_s}\right)^2},
\label{dens.dm}
\end{equation}
we obtain the values $\rho_s=6.14\times10^3 \rho_c$ and $r_s=378.5$ kpc, where $\rho_c$ is the critical density of the universe.

\begin{figure}
\begin{center}
{
 \epsfig{file=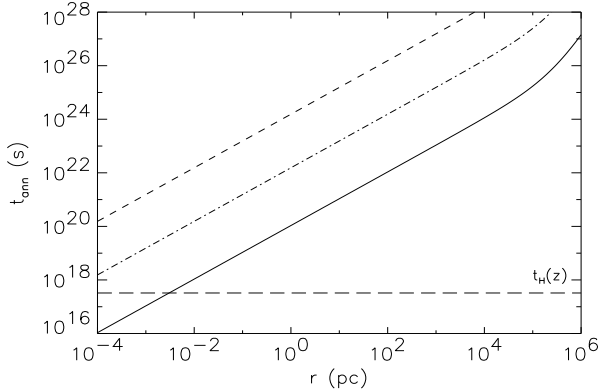,height=5.7cm,width=8.5cm,angle=0.0}
}
\end{center}
 \caption{Annihilation time (as defined in eq.\ref{ann.time}) for a DM halo with $M_\chi=500$ GeV
and the density parameters of the DME region for $\langle \sigma v \rangle=4.3\times10^{-22}$ cm$^3$ s$^{-1}$ (solid line), $\langle \sigma v \rangle=3\times10^{-26}$ cm$^3$ s$^{-1}$ (dashed line) and $\langle \sigma v \rangle=3\times10^{-24}$ cm$^3$ s$^{-1}$ (dot-dashed line), compared with the Hubble time at the Bullet Cluster redshift (long dashed line).}
 \label{tann_bullet_b3}
\end{figure}

Formally, this DM density profile tends to infinite value at $r\rightarrow 0 $; indeed, this is not physically possible because the increasing DM density at the cluster center increases the annihilation probability, and therefore the annihilation time scale
\begin{equation}
t_{ann}(r)=\frac{1}{n_{DM}(r) \langle \sigma v \rangle} ,
\label{ann.time}
\end{equation}
where $n_{DM}(r)=\rho(r)/M_\chi$ is the neutralino numerical density, and $\langle \sigma v \rangle$ is the mean value of the annihilation cross section, decreases. 
When the annihilation time given in eq.(\ref{ann.time}) is lower than the Hubble time $t_H$ at the cluster redshift, the annihilation processes are dominating and, as a consequence, the DM density cannot increase again. So, one can find a minimum radius over which the DM distribution has the shape of eq.(\ref{dens.dm}), and under which the density is roughly constant. 
In Figure \ref{tann_bullet_b3} we show the annihilation time for a neutralino mass of $M_\chi=500$ GeV and for three values of the cross section:
a reference value of $\langle \sigma v \rangle=3\times10^{-26}$ cm$^3$ s$^{-1}$ (this is an estimate of the DM relic abundance value), a value of the order of the cross section we will find in next analysis, $\langle \sigma v \rangle=4.3\times10^{-22}$ cm$^3$ s$^{-1}$, and an intermediate value of $\langle \sigma v \rangle=3\times10^{-24}$ cm$^3$ s$^{-1}$. 
As we can see, the radius at which the DM density stops raising is quite small; for a cross section value of $\langle \sigma v \rangle=4.3\times10^{-22}$ cm$^3$ s$^{-1}$ it is $\sim3\times10^{-3}$ pc, and it is smaller for decreasing cross section values.

By using this radial distribution density for DM, we calculate the production rate of secondary electrons and the equilibrium spectrum due to energy losses following the procedure described in Colafrancesco et al. (2006). In this calculation, we assume different masses and compositions for the neutralino by using the DarkSusy package (Gondolo et al. 2004). We assume here a boosting factor due to the subhalos DM distribution (Pieri et al. 2011) equal to 1, and larger values can be considered eventually simply by multiplying the result by this factor (see, e.g., Colafrancesco et al. 2011b).

\begin{figure}
\begin{center}
{
 \epsfig{file=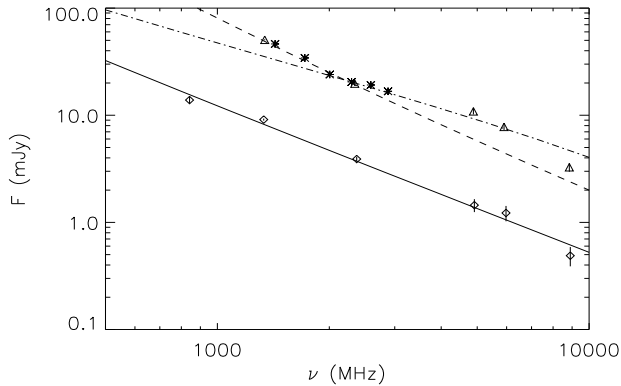,height=5.7cm,width=8.5cm,angle=0.0}
}
\end{center}
 \caption{Radio emission in the different regions of the Bullet cluster with:
a WR model with $s_p=2.7$ and $B=7.5$ $\mu$G in the MS (solid line),
a WR model with $s_p=2.9$ and $B=60$ $\mu$G in the BS (dashed line) and
a DM model with $M_\chi=500$ GeV, composition $W^+W^-$, $B=0.015$ $\mu$G and $\langle \sigma v \rangle=8.1\times10^{-17}$ cm$^3$ s$^{-1}$ in the DME region.
Data are taken from region \# 3 of Liang et al. (2000) (diamonds),
from region \# 2 of Liang et al. (2000) (triangles)
and from the smaller region of Shimwell et al. (2014) (asterisks).}
 \label{radio_3comp}
\end{figure}

\begin{figure}
\begin{center}
{
 \epsfig{file=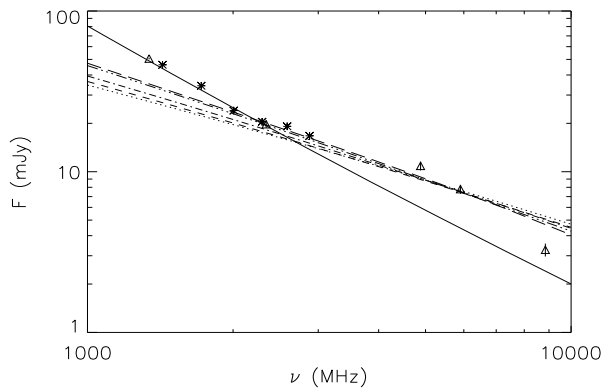,height=5.7cm,width=8.5cm,angle=0.0}
}
\end{center}
 \caption{Radio emission in the different regions of the Bullet cluster with: 
a WR model with $s_p=2.9$ and $B=60$ $\mu$G in the BS (solid line) and 
several DM models with $M_\chi=500$ GeV and composition $W^+W^-$ in the DME region with 
$B=0.015$ $\mu$G and $\langle \sigma v \rangle=8.1\times10^{-17}$ cm$^3$ s$^{-1}$ (long dashed line),
$B=1$ $\mu$G and $\langle \sigma v \rangle=1.6\times10^{-20}$ cm$^3$ s$^{-1}$ (three dotted-dashed line),
$B=5$ $\mu$G and $\langle \sigma v \rangle=1.1\times10^{-21}$ cm$^3$ s$^{-1}$ (dot-dashed line),
$B=10$ $\mu$G and $\langle \sigma v \rangle=6.5\times10^{-22}$ cm$^3$ s$^{-1}$ (dashed line),
$B=20$ $\mu$G and $\langle \sigma v \rangle=5.8\times10^{-22}$ cm$^3$ s$^{-1}$ (dotted line).
Data are taken from region \# 2 of Liang et al. (2000) (triangles)
and from the smaller region of Shimwell et al. (2014) (asterisks).}
 \label{radio_high_DM}
\end{figure}

First, we check if the spectral shape of the radio emission from DME can explain the flattening of the radio spectrum at $\nu\ge2.3$ GHz. 
In Figure \ref{radio_3comp} we show the emissions produced using the WR model in the MS and BS regions, using for the BS the model with $s_p=2.9$, that reproduces well the data up to 2.3 GHz, and we show the radio spectrum produced in the DME region for $M_\chi=500$ GeV, composition $W^+W^-$, $B=0.015$ $\mu$G and $\langle \sigma v \rangle=8.1\times10^{-17}$ cm$^3$ s$^{-1}$. 
We discuss this model because it produces the best result, regarding to the spectral shape of radio flux w.r.t. the observed one, among all the models we considered, for neutralino masses of $M_\chi=9$, 60 and 500 GeV, compositions $b \bar b$, $\tau^+ \tau^-$ and $W^+ W^-$, and by searching for the value of the magnetic field that produces the spectral shape closest to the observed one, and calculating the corresponding cross section to fit the radio data.
The radio emission from DME is subdominant for $\nu \le 2.3$ GHz, and is dominant at higher frequencies reproducing the correct slope up to 5.9 GHz, whereas it overestimates the point at 8.8 GHz. In addition to this problem, this model is clearly unrealistic, because it requires a value of the magnetic field smaller than the lower limit of 0.2 $\mu$G found with HXR measurements by Wik et al. (2014), and a DM annihilation cross section  larger by a factor of $\sim10^8$ w.r.t. the upper limit $\langle \sigma v \rangle \sim10^{-24}$ cm$^3$ s$^{-1}$ obtained by the Fermi analysis of dwarf galaxies (Ackermann et al. 2014).
Anyway, it is possible to obtain a similar spectrum by changing the magnetic field value and the corresponding cross section (see Figure \ref{radio_high_DM}). 
For higher values of the magnetic field, the resulting spectrum shows a smaller curvature at high frequencies, so it overestimates the point at 8.8 GHz by a larger factor w.r.t. the previous model, but it is compatible with the HXR measures for $B \ge 0.055$ $\mu$G and, for $B\simgt 5$ $\mu$G, it requires  a cross section higher than the Fermi upper limits by a factor less than $10^3$, that can be recovered by assuming the presence of a boosting factor originated by the substructures in the halo.
We stress, however, that the problem of the spectral steepening at 8.8 GHz is present in each of these models; we will discuss this problem in Sect. 3.1 below.

\begin{figure}
\begin{center}
{
 \epsfig{file=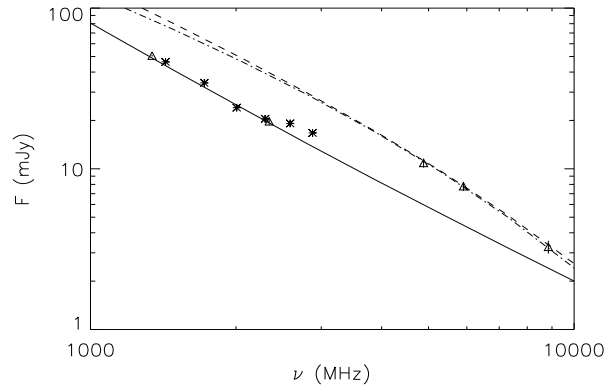,height=5.7cm,width=8.5cm,angle=0.0}
}
\end{center}
 \caption{Radio emission in the different regions of the Bullet cluster with: 
a WR model with $s_p=2.9$ and $B=60$ $\mu$G in the BS (solid line) and 
several DM models with
$M_\chi=9$ GeV in the DME region, with composition $b \bar b$, 
$B=30$ $\mu$G and $\langle \sigma v \rangle=4.6\times10^{-23}$ cm$^3$ s$^{-1}$ (dashed line),
and composition $\tau^+ \tau^-$, 
$B=10$ $\mu$G and $\langle \sigma v \rangle=2.6\times10^{-23}$ cm$^3$ s$^{-1}$ (dot-dashed line).
Data are taken from region \# 2 of Liang et al. (2000) (triangles)
and from the smaller region of Shimwell et al. (2014) (asterisks).}
 \label{radio_high_DM9}
\end{figure}

For the model with $M_\chi=500$ GeV and composition $W^+ W^-$, the gamma ray flux produced in the case with $B=1$ $\mu$G is $1.8\times10^{-11}$ cm$^{-2}$ s$^{-1}$, and for $B=0.015$ $\mu$G it is $9.3\times10^{-8}$ cm$^{-2}$ s$^{-1}$ (greater than the Fermi upper limit in the last case).

For comparison, we show also the results for two models with $M_\chi=9$ GeV (see Figure \ref{radio_high_DM9}): they reproduce well the observed steepening at $\nu \simgt 5$ GHz, and provide ICS fluxes of $4.4\times10^{-18}$ erg cm$^{-2}$ s$^{-1}$ for the $b \bar b$ model, and $8.0\times10^{-17}$ erg cm$^{-2}$ s$^{-1}$ for the $\tau^+ \tau^-$ model, both smaller than the observed upper limit of Wik et al. (2014). However, these models overestimate the radio data at $\nu \le 3$ GHz.

\subsection{The Dark Matter Western peak}

Now we calculate the radio emission produced by DM models in the DMW region, where a strong radio component does not seem to be present.

From gravitational lensing results, the DMW halo is less massive than the DME, with a mass of $M_{DMW}\sim3.0\times10^{14} M_\odot$ inside 310 kpc (Bradac et al. 2006). The NFW parameters we derive are $\rho_s=7.41\times10^3 \rho_c$ and $r_s=283.5$ kpc.

For the model with $M_\chi=500$ GeV and composition $W^+ W^-$ we find that, for the same values of the magnetic field and the cross section used for the DME, the radio emission produced in the DMW is smaller by a factor of $\sim 0.6$ than the emission in DME. So, this emission is relatively low, but in principle not negligible, considering also that in this region, near to the shock front, strong magnetic fields and recently accelerated relativistic particles can be present. We will discuss this problem in Section 3.2 below.

\section{Refining and testing the basic model}

\subsection{High frequencies spectrum}

We discuss here the problem related to the excess flux at 8.8 GHz, that is difficult to be explained in all the models we considered. We consider different possible explanations:\\
\textit{i)} our model is simplified because it takes into account only the radio emissions coming from the three described regions, while the region \# 2 in Liang et al. (2000) contains the contributions from a larger area. Other contributions can flatten the total emission between 2.3 and 5 GHz, and steepen it at 8.8 GHz;\\
\textit{ii)} the experimental technique to obtain the data is affected by some issue: for example, the points of Liang et al. (2000) are corrected for the thermal SZE expected at the same frequencies, and these corrections are calculated under the assumption of an isothermal cluster with the density given by two beta models centered on the two X-rays peaks, following Andreani et al. (1999). However, the assumption of isothermality can be too inaccurate in this very complex cluster (e.g., Markevitch et al. 2002, Prokhorov \& Colafrancesco 2012), and also other SZE components can be present, and they can affect the total flux in different ways at different frequencies (see Colafrancesco et al. 2011a);\\
\textit{iii)} an incorrect source subtraction, which is a delicate operation mostly for extended sources, can alter the total radio halo spectrum.
Both Liang et al. (2000) and Shimwell et al. (2014) report a detailed list of sources in the field. Among the listed sources contained inside the region \# 2 of Liang et al. (2000), the most interesting one appears to be the source labeled with an A by Liang et al. (2000) and with an L by Shimwell et al. (2014), at coordinates (J2000) RA 06:58:37.9 and Dec -55:57:25. 
This is the only source inside this field with a quite high flux ($\sim20$ mJy at 1.3 GHz), and it is slightly extended compared with the ATCA resolution of 2.7" in Shimwell et al. (2014), 
so the procedure of the flux removal can induce some errors in the total flux estimate of the halo.

The A(L) radio source is located near the galaxy 2MASX J06583806-5557256, in a region rich of galaxies, at the coordinates J2000 RA 06:58:38.0 and Dec -55:57:26. It is detected in the infrared bands by WISE and in the 2MASS survey with $K_s$, $H$ and $J$ magnitudes 13.4, 14.0 and 15.0, respectively, and in the optical band with B magnitude 20.3 and redshift $z=0.297$ (Barrena et al. 2002). Another galaxy that is very close the previous one is located at coordinates J2000 RA 06:58:38.0 and Dec -55:57:23, with B magnitude 21.0 and redshift $z=0.292$ (Barrena et al. 2002). The position of the A(L) source, located close (and likely corresponding) to the SUMSS source at coordinates J2000 RA 06:58:37.6 and Dec -55:57:24, is between these two galaxies, and it is not clear if it is associated with one of them, or it is related to a possible interaction.

We also note that the position of the radio source A(L) is very close to the center of the DME region.
This fact suggests also the possibility that the radio source A(L) can be produced by the DME halo, in the same way we explored in the previous Section. Since for a NFW DM density profile the radio emission has a sharp central peak and an extended halo component, it is possible that this source is the central emission of the DME region that has been subtracted (that at the distance of the Bullet cluster has a radius of few arcsec). In this case, to reproduce the total emission from DME, the DM model should be normalized to the A(L) source spectrum, and its emission should be added to the total halo emission.

The spectrum of the A(L) source can be reproduced with a DM model with $M_\chi=500$ GeV,  $W^+ W^-$ composition, $B=10$ $\mu$G and $\langle \sigma v \rangle=4.3\times10^{-22}$ cm$^3$ s$^{-1}$. 
The same flux level can be obtained also with a cross section of the order of the maximum value allowed by the Fermi upper limit from Dwarf Galaxies ($\langle \sigma v \rangle \sim 10^{-24}$ cm$^3$ s$^{-1}$, Ackermann et al. 2014) and a boosting factor smaller than $10^3$.
We found that models with other values of the neutralino mass or different compositions fail to reproduce the correct slope of the radio flux, but we note that there is a degeneracy between the value of the magnetic field and that of the cross section (Colafrancesco et al. 2006, Colafrancesco et al. 2015).

\begin{figure}
\begin{center}
{
 \epsfig{file=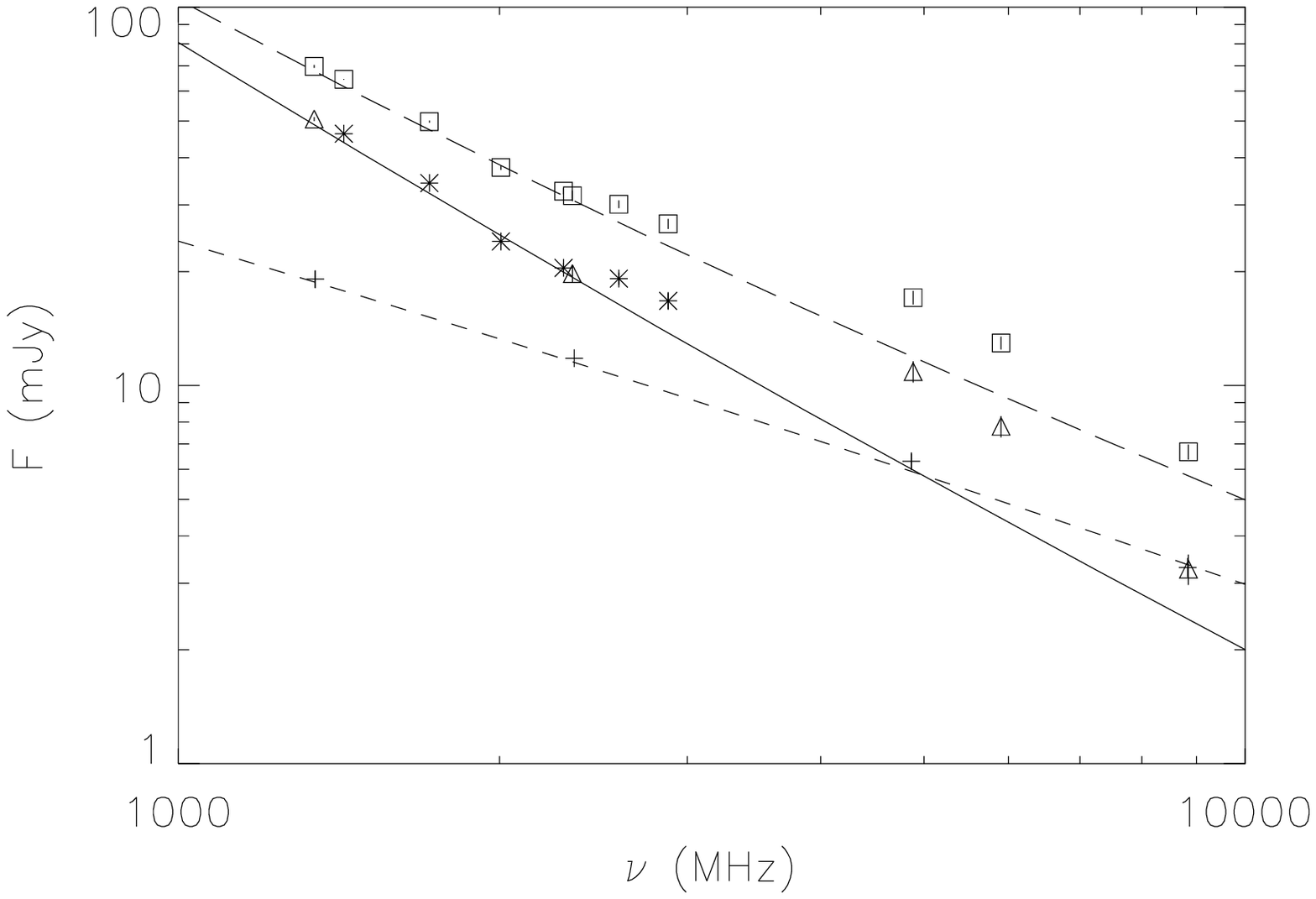,height=5.7cm,width=8.5cm,angle=0.0}
 \epsfig{file=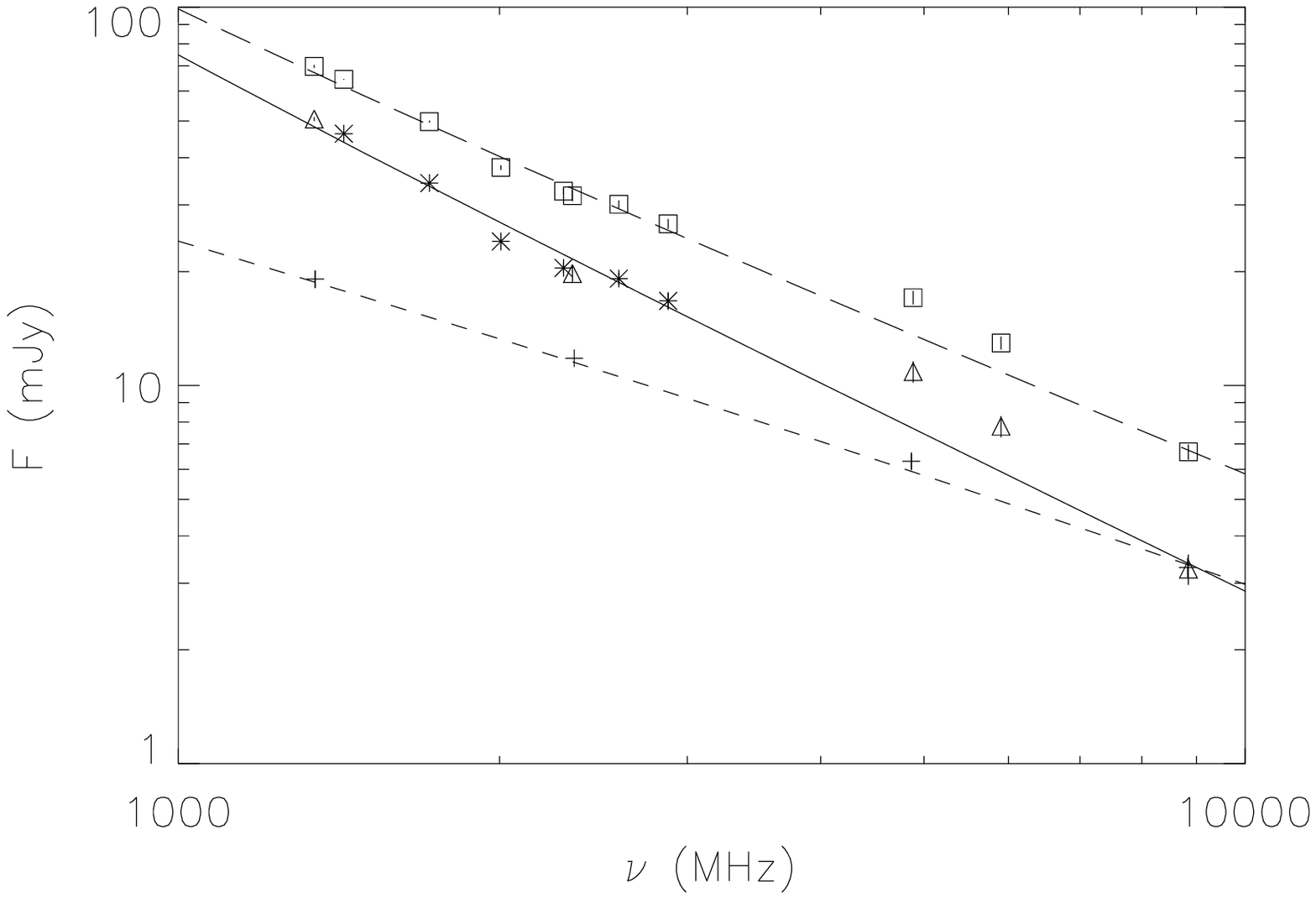,height=5.7cm,width=8.5cm,angle=0.0}
}
\end{center}
 \caption{Upper panel: radio emission from different regions in the Bullet cluster with
a WR model with $s_p=2.9$ and $B=60$ $\mu$G in the BS (solid line) and 
a DM model with $M_\chi=500$ GeV, composition $W^+ W^-$, $B=10$ $\mu$G and $\langle \sigma v \rangle=4.3\times10^{-22}$ cm$^3$ s$^{-1}$ (dashed line) in the DME region, and their sum (long dashed line).
Data are from region \# 2 of Liang et al. (2000) (triangles), from the smaller region of Shimwell et al. (2014) (asterisks), and from radio source A(L) from Liang et al. (2000) (crosses).
With the squares we plot the sum of the cluster data and the radio source date at the same frequencies (derived with interpolation at the frequencies where they are not available in Liang et al. 2000).
Lower panel: as in the upper panel, but for a WR model in the BS with $s_p=2.7$ and $B=27$ $\mu$G.}
 \label{radio_c2_3_wr+dm}
\end{figure}

In Fig.\ref{radio_c2_3_wr+dm} we plot the radio spectrum from the region \# 2, reproduced with the two WR models with different $s_p$ values we discussed before in the BS, compared with the data from this region (triangles and asterisks), and the emission of the radio source A(L) explained with the DM model with the corresponding data (crosses). In the same figure we also show the sum of the models in the two regions compared with the sum of the data of the cluster and the radio source (squares); we stress that in order to obtain these data at the frequencies not included in the Table 2 of Liang et al. (2000) we interpolated the data with a power-law model. As we can see, the best case seems to take place for the WR model in the BS with $s_p=2.7$, even if it underestimates the data at 5 and 6 GHz. The corresponding magnetic field value is 27 $\mu$G; this is a quite high value, but we notice that the BS is located near the shock front, and that usually the magnetic fields in cool core clusters are stronger than those in other clusters (e.g., Carilli \& Taylor 2002). 

The model with $M_\chi=500$ GeV, $W^+ W^-$ composition, magnetic field 10 $\mu$G and cross section $\langle \sigma v \rangle=4.3\times10^{-22}$ cm$^3$ s$^{-1}$ produces an ICS flux in the 50--100 keV band of $1.9\times10^{-17}$ erg cm$^{-2}$ s$^{-1}$, smaller than the upper limit derived by Wik et al. (2014).

The gamma ray flux for $E>0.1$ GeV predicted by the same model is $6.3\times10^{-13}$ cm$^{-2}$ s$^{-1}$ (given by the sum of $\pi^0$ decay and ICS and non-thermal bremsstrahlung from electrons), smaller than the limit derived by Ackermann et al. (2010).

\subsection{Emission from the DMW region}

\begin{figure}
\begin{center}
{
 \epsfig{file=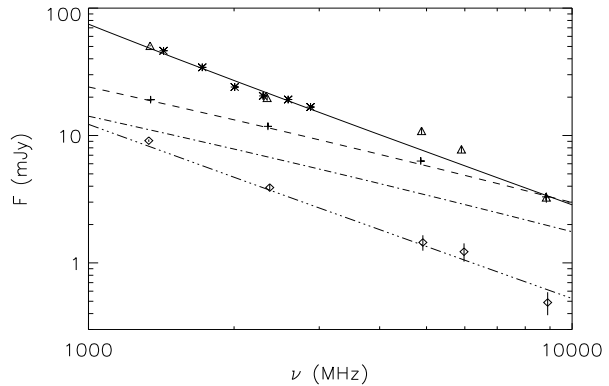,height=5.7cm,width=8.5cm,angle=0.0}
}
\end{center}
 \caption{Radio emission from different regions in the Bullet cluster with
a WR model with $s_p=2.7$ and $B=27$ $\mu$G in the BS (solid line), 
two DM models with $M_\chi=500$ GeV, composition $W^+ W^-$, $B=10$ $\mu$G and $\langle \sigma v \rangle=4.3\times10^{-22}$ cm$^3$ s$^{-1}$ in the DME region (dashed line) and in the DMW region (dot-dashed line), and a WR model with $s_p=2.7$ and $B=7.5$ $\mu$G in the MS (three dots-dashed line).
Data are from region \# 2 of Liang et al. (2000) (triangles), from the smaller region of Shimwell et al. (2014) (asterisks), from radio source A(L) from Liang et al. (2000) (crosses), and region \# 3 of Liang et al. (2000) (diamonds).}
 \label{radio_4comp}
\end{figure}

By considering the model we discussed previously with i) a WR model with $s_p=2.7$ and $B=27$ $\mu$G in the BS to explain the spectrum in the region \# 2, ii) a DM model with $M_\chi=500$ GeV, composition $W^+ W^-$, $B=10$ $\mu$G and $\langle \sigma v \rangle=4.3\times10^{-22}$ cm$^3$ s$^{-1}$ in the DME region to explain the spectrum of the A(L) radio source, and iii) a WR model with $s_p=2.7$ and $B=7.5$ $\mu$G in the MS to explain the emission in the internal region, we consider the contribution of a DM model coming from the DMW region, by taking in account the smaller mass of this halo and with the same properties of the neutralino (mass, composition and annihilation cross section) used in the DME model.
The spectrum resulting from this model is plotted in the Fig.\ref{radio_4comp}. The emission from the DMW region is lower than the emission from the DME, but it is not totally negligible. This might be a problem because the radio map does not show a strong emission in this region; there are some radio sources located in this region (the A, D, and E sources in Shimwell et al. 2014; in particular the A source is located very near to the DMW peak), but their flux is small. Also, in this plot we assumed that the magnetic field is the same in the DME and DMW regions, while in the DMW we can expect that it is higher than in the DME because the DMW region is located near the shock.
However, it is possible that the magnetic field can be amplified by the shock compression in the region to the west of the shock front, while it can be reduced in the region to the east of the shock front, where the DMW is located, because of the post-shock dilution.\\
Therefore, it seems necessary to perform a deeper study of the radio emission from this region to better understand the properties of the DMW halo.

\begin{figure}
\begin{center}
{
 \epsfig{file=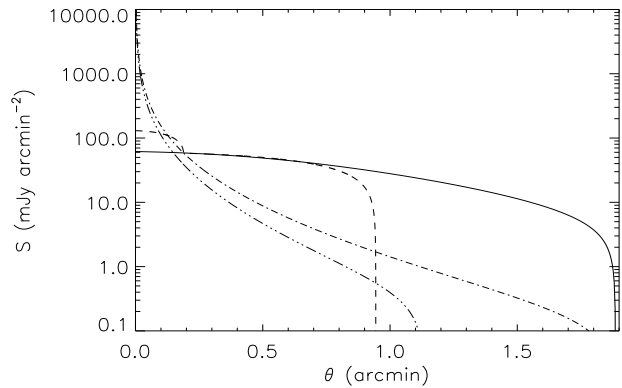,height=5.7cm,width=8.5cm,angle=0.0}
}
\end{center}
 \caption{Surface brightness profiles at 1.5 GHz for:
  a WR model with $s_p=2.7$ and $B=7.5$ $\mu$G in the MS (solid line),
  a WR model with $s_p=2.7$ and $B=27$ $\mu$G in the BS (dashed line),
  two DM models with $M_\chi=500$ GeV, composition $W^+ W^-$, $B=10$ $\mu$G and $\langle \sigma v \rangle=4.3\times10^{-22}$ cm$^3$ s$^{-1}$ in the DME region (dot-dashed line) and in the DMW region (three dots-dashed line).}
 \label{bril_4comp}
\end{figure}

To compare the combined effect of the different components with the observed radio map, we calculate the surface brightness profiles at 1.5 GHz for the different models shown in the Fig.\ref{radio_4comp}. We plot these profiles in  Fig.\ref{bril_4comp}, centering all the profiles on the same point. As we can see, the surface brightness produced in the DMW region inside 0.1 arcmin from the center is not negligible, and at the distance of $\sim1$ arcmin it is of the order of 0.5 mJy arcmin$^{-2}$, so it should be detected with SKA if integrated in a region of a few arcsec size, provided that the sensitivity of SKA at this frequency is of the order of $\sim \mu$Jy (Dewdney et al. 2012).

We show also a simulated radio map of the surface brightness profiles plotted in the Fig.\ref{bril_4comp}, centering the profiles on the corresponding positions on the map. The result is shown in the first panel of the Fig.\ref{mappe_bullet_varie}. This map can be compared to the observed radio map reported in Fig.5 of Shimwell et al. (2014). In the simulated map the DM emissions have narrow and intense peaks, but also extended residual emissions; so, it is possible that the peaks have been interpreted as point-like (or slightly extended) sources, while the residual halo is included in the overall cluster emission. This can be the case for the DME, while, as discussed before, it seems to be a more difficult expectation for the DMW.
We also note that the emission produced in the BS, in addition to have a weaker central emission w.r.t. the DM peaks, has a limited spatial extension, that does not explain all the extended emission observed, especially on the north and on the south of this region. So, to describe the full size of the radio emission observed in this region, it is probably necessary to calculate the emission in a wider region surrounding the BS, or to consider the contributions coming from other sources, like the galaxies halos.

\subsection{X-ray and gamma-ray emission}

\begin{figure}
\begin{center}
{
 \epsfig{file=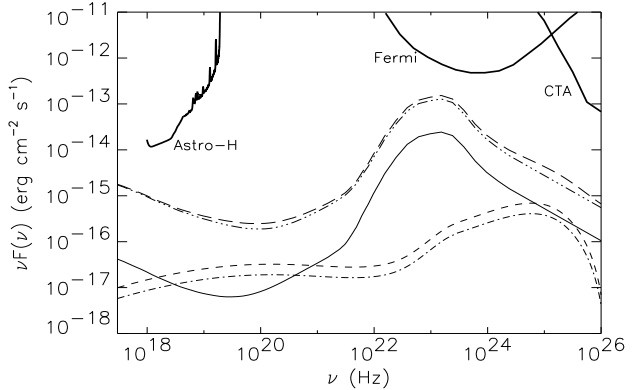,height=5.7cm,width=8.5cm,angle=0.0}
}
\end{center}
 \caption{X-rays and gamma rays emission in the different regions of the Bullet cluster with:
   a WR model with $s_p=2.7$ and $B=7.5$ $\mu$G in the MS (three dots-dashed line),
  a WR model with $s_p=2.7$ and $B=27$ $\mu$G in the BS (solid line),
  two DM models with $M_\chi=500$ GeV, composition $W^+ W^-$, $B=10$ $\mu$G and $\langle \sigma v \rangle=4.3\times10^{-22}$ cm$^3$ s$^{-1}$ in the DME region (dashed line) and in the DMW region (dot-dashed line).
The long dashed line is the sum of all the components. 
We show also the sensitivity curves for Astro-H with 100 ks of time integration (from http://astro-h.isas.jaxa.jp/researchers/sim/sensitivity.html), Fermi-LAT for 10 yrs, and CTA for 1000 hrs (from Funk \& Hinton 2013).}
 \label{xray_gamma_4comp}
\end{figure}

In Fig.\ref{xray_gamma_4comp} we show the X-ray and gamma-ray fluxes produced in the four regions considered for the models shown in the Fig.\ref{radio_4comp}. These emissions are the sum of ICS, non-thermal bremsstrahlung and $\pi^0$ decay.
As we can see in this figure, we expect a main peak of the spectral energy distribution (SED) around $\nu\sim10^{23}$ Hz coming from hadronic models (i.e. the WR model) and a lower one around $\nu\sim10^{25}$ Hz coming from DM annihilation models. The model is dominated by the emission from the MS.

The calculated fluxes are lower than the current upper limits ($9.1\times10^{-13}$ erg cm$^{-2}$ s$^{-1}$ at $\nu\sim2.4\times10^{22}$ Hz by Fermi, and $1.1\times10^{-12}$ erg cm$^{-2}$ s$^{-1}$ at $\nu\sim(1.2\div2.4)\times10^{19}$ Hz by NuSTAR). We also show for the sake of comparison the sensitivity curves for Astro-H with 100 ks of time integration (from http://astro-h.isas.jaxa.jp/researchers/sim/sensitivity.html), Fermi-LAT for 10 yrs, and CTA for 1000 hrs (from Funk \& Hinton 2013). It is clear that the high energy emission is not expected to be detected with these instruments, due to the relatively large distance of this cluster.

We conclude that the main source of information about the origin of DM and non-thermal plasmas in the Bullet cluster, also by considering the next generation of instruments, is the radio band where the very high sensitivity of instruments like the SKA can be used to probe the physics of this DM-dominated system.
Note that the SKA will have also the possibility to measure simultaneously the magnetic field in the various regions of the Bullet cluster using quite precise Faraday rotation measurements (see Johnston-Hollitt et al. 2014) and thus it will be able to fully disentangle the degeneracy between DM-produced secondary electrons and magnetic field in the various sub halos of this cluster.\\
Another possibility that we discuss in next section is to use the SZE, by using the fact that it does not depend on the distance of the cluster, so it is suitable to study the Bullet cluster. We notice that another important property of the SZE is that it does not depend on the magnetic field, so it can allow to break the degeneracy between electrons density and magnetic field intensity existing in the radio continuum synchrotron emission.

\section{The Sunyaev-Zel'dovich effect in the Bullet cluster}

The SZE is produced by the ICS of CMB photons off the electrons contained in ionized halos of galaxy clusters and other cosmic structures (Zel'dovich \& Sunyaev 1969, Sunyaev \& Zel'dovich 1972). The SZE can be calculated for both thermal or non-thermal halos when the full relativistic formalism is used (see, e.g., Colafrancesco et al. 2003 for details).\\
The SZE from the Bullet cluster has been already studied in Colafrancesco et al. (2011a) by using its spectral properties over a wide frequency range from 18 GHz to 850 GHz. Here we perform a more detailed analysis regarding both the spectral and spatial properties, and using the results obtained in previous sections to constrain the properties of the non-relativistic electrons in the Bullet cluster.

\subsection{Spectral analysis}

The SZE in the Bullet cluster has been measured over a wide range of frequencies, with ACBAR at 150 and 275 GHz (Gomez et al. 2004), with the SEST telescope at 150 GHz (Andreani et al. 1999), with APEX at 150 GHz (Halverson et al. 2009), with the SPT at 150 GHz (Plagge et al. 2010), with ATCA at 18 GHz  (Malu et al. 2010), and with Herschel-SPIRE at 600, 850 and 1200 GHz (Zemcov et al. 2010).
The combination of the measurements from low to high frequencies allowed us to obtain information on the presence of multiple components in the overall SZE signal, favoring the presence of a second component of non-thermal origin, although a second thermal component with very high temperature cannot be excluded (Colafrancesco et al. 2011a).

Since the SZE signal seems to be located very close to the MS region (see next subsection for a detailed discussion), we perform here a more refined spectral analysis, by taking into account the properties of the thermal and non-thermal components in the MS as discussed in the previous Sections.

\begin{figure}
\begin{center}
{
 \epsfig{file=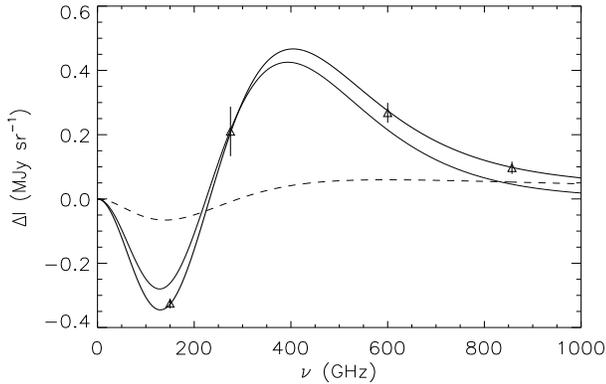,height=5.7cm,width=8.5cm,angle=0.0}
}
\end{center}
 \caption{SZE in the Bullet cluster, with the thermal effect in the MS with $kT=14.2$ keV and $\tau=1.1\times10^{-2}$ (solid line), a non-thermal effect with $s=3.7$, $p_1=1$ and $\tau=3\times10^{-4}$ (dashed line), and the sum of the two components (thick line).
 Data at 150 and 275 GHz are taken with ACBAR (Gomez et al. 2004), and data at 600 and 857 GHz with Herschel-SPIRE (Zemcov et al. 2010).}
 \label{sz_2pop_nont}
\end{figure}

\begin{figure}
\begin{center}
{
 \epsfig{file=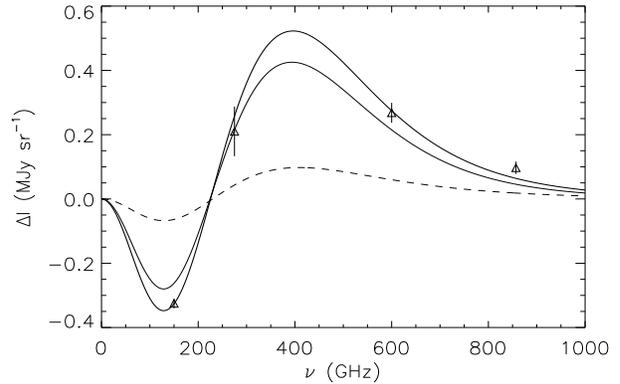,height=5.7cm,width=8.5cm,angle=0.0}
}
\end{center}
 \caption{SZE in the Bullet cluster, with the thermal effect in the MS with $kT=14.2$ keV and $\tau=1.1\times10^{-2}$ (solid line), a  second thermal effect with $kT=25$ keV and $\tau=1.5\times10^{-3}$ (dashed line), and the sum of the two components (thick line).
 Data at 150 and 275 GHz are taken with ACBAR (Gomez et al. 2004), and data at 600 and 857 GHz with Herschel-SPIRE (Zemcov et al. 2010).}
 \label{sz_2pop_therm}
\end{figure}

In Colafrancesco et al. (2011a), the two best possibilities to explain the SZE data at all the frequencies were: \textit{i)} a thermal component with $kT=13.9$ keV and optical depth $\tau=1.1\times10^{-2}$ and a non-thermal component with a power law electrons spectrum with spectral index $s=2.7$, normalized minimum momentum $p_1=1$ and $\tau=2.3\times10^{-4}$; \textit{ii)} a thermal component with $kT=13.9$ keV and $\tau=3.5\times10^{-3}$ and a second thermal component with $kT=25$ keV and $\tau=5.5\times10^{-3}$. The first possibility was slightly favoured w.r.t. the second one.\\
Now we can perform a more accurate analysis by setting the parameters of the first thermal component to the values assumed for the MS, corresponding to a temperature of 14.2 keV (Wik et al. 2014) and an optical depth of $\tau=1.1\times10^{-2}$ (calculated with the parameters of Ota \& Mitsuda 2004), and considering the two cases of a second component of non-thermal or thermal origin. For the non-thermal case, we set $s=3.7$, from the value of the radio spectrum in the MS (region \# 3 in Liang et al. 2000), $p_1=1$ (from Colafrancesco et al. 2011a), and leaving $\tau$ as a free parameter; for the second thermal component case we set $kT=25$ keV (from Colafrancesco et al. 2011a), and leave $\tau$ as a free parameter.
The results are reported in the Fig.\ref{sz_2pop_nont} for the non-thermal case, and in the Fig.\ref{sz_2pop_therm} for the two thermal components case. As we can see, the case with a non-thermal component provides a better fit to the data than the case with two thermal component, especially considering the point at 857 GHz, hence confirming the results obtained by Colafrancesco et al. (2011a).

\begin{figure}
\begin{center}
{
 \epsfig{file=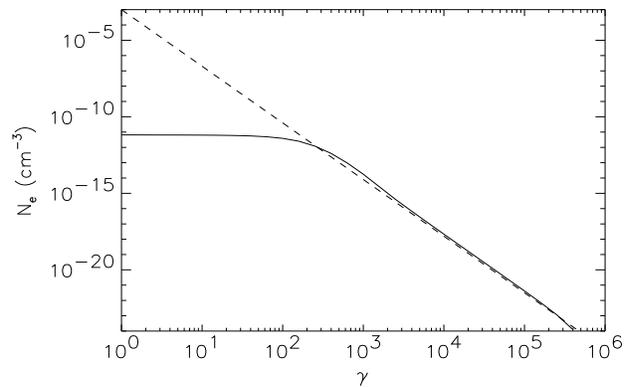,height=5.7cm,width=8.5cm,angle=0.0}
}
\end{center}
 \caption{Secondary electrons spectrum derived in the WR model in the MS (solid line) compared with the primary electrons spectrum derived from the SZE spectral study (dashed line).}
 \label{elettr_b1_prim_sec}
\end{figure}

It is interesting to compare the non-thermal component resulting from the analysis of the SZE spectrum with the electrons coming from the WR model in the MS. From the SZE we derived a very good agreement with the data by using $p_1=1$, so a very low value of the minimum momentum of the electron distribution is required. In the secondary electron models (of both hadronic and DM origin) the spectrum of the electrons flattens for $p\simlt 10^2$, because of the Coulombian losses effect and of the intrinsic flattening of the source term (see, e.g., Marchegiani et al. 2007 for the hadronic case, and Colafrancesco et al. 2006 for the DM case), and this fact strongly suppresses the presence of electrons with small momentum $p$. So, these results point towards a different origin of the electrons, i.e. ``primary" electrons, recently accelerated by some mechanism (possibly connected to the recent merging of the two subclusters).
If we assume that these electrons have the same spatial distribution of the thermal gas, and if we write their spectrum as $N_e(\gamma)=k_0 \gamma^{-s}$ with $s=3.7$, from the value of the optical depth derived from the SZE analysis we obtain a normalization factor of $k_0=1.0\times10^{-3}$ cm$^{-3}$.
In Fig.\ref{elettr_b1_prim_sec} we compare this spectrum to the spectrum of the secondary electrons derived from the WR model in the MS. It is interesting to note that the two spectra overlap almost exactly for $\gamma \simgt 10^3$, i.e. in the energy region of the electrons that produce the radio emission. 
This means that all the considerations we made for the WR model in the MS hold also for this model (actually, a slightly higher value of the magnetic field, $B\sim8$ $\mu$G, is required to fit the radio halo), with the difference that in this model the contribution of the $\pi^0$ decay to the gamma-ray emission is not present, and therefore a much lower gamma-ray emission (produced only by non-thermal bremsstrahlung and ICS processes) is expected. We can then conclude that radio measurements alone  cannot discriminate between the primary and secondary models, while the SZE observations indicate that a primary origin is favoured.

\subsection{Spatial analysis}

There is some difficulty in the comparison between different maps of the SZE obtained so far for the Bullet Cluster: in fact, while Halverson et al. (2009) at 150 GHz and Zecomv et al. (2010) at 600 GHz found that the peak of the SZE emission is located near the MS, a bit elongated towards the direction of the BS, Malu et al. (2010) at 18 GHz and Plagge et al. (2010) at 150 GHz found that the peak of the SZE is shifted from the MS in the direction of the DME.

\begin{figure}
\begin{center}
{
 \epsfig{file=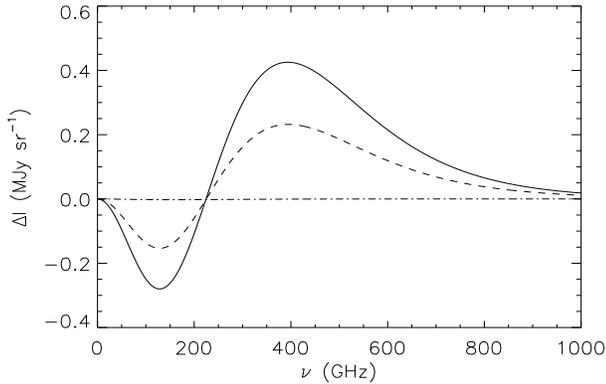,height=5.7cm,width=8.5cm,angle=0.0}
}
\end{center}
 \caption{SZE in the Bullet Cluster produced by different components:
 a thermal effect in MS (solid line), a thermal effect in BS (dashed line) and a DM model in the DME region (dot-dashed line), that is the dominant component among the non-thermal effects in the other regions. The experimental data are not shown here because no one of these models alone is able to reproduce them (see Figs.\ref{sz_2pop_nont} and \ref{sz_2pop_therm}).}
 \label{sz_4comp}
\end{figure}

To compare our results with these maps, we calculate the SZE produced in the different regions of the cluster for the different models used in the previous sections. Therefore, we calculate the thermal effects in the MS and BS, the effects for the WR models in MS and BS, and the effects for the DM models in DME and DMW regions (using the DME model normalized to the A(L) source), by using the full relativistic formalism (Colafrancesco et al. 2003) for all the models, and the formalism for the cool-core clusters for the BS (Colafrancesco \& Marchegiani 2010). 
First, we compare the different curves so far obtained by putting them in the same plot (see Fig.\ref{sz_4comp}). We find that the thermal effects are dominant, and the effect in the MS is dominant w.r.t. the effect in the BS, while the non-thermal effects by secondary electrons (both of hadronic and DM origin) are negligible, hence confirming that a primary electron population is required in order to reproduce the experimental data.

\begin{figure}
\begin{center}
{
 \epsfig{file=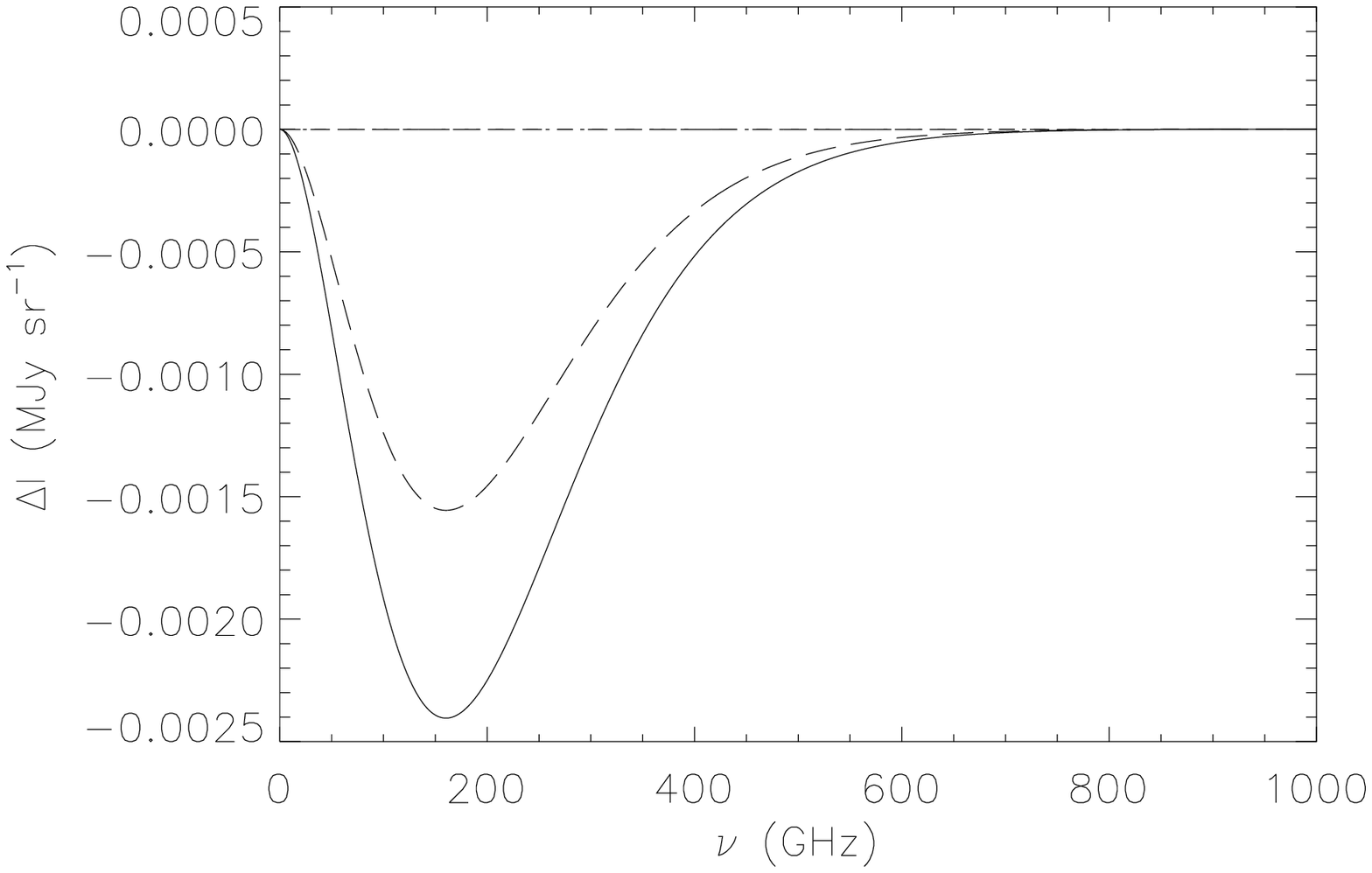,height=5.7cm,width=8.5cm,angle=0.0}
 \epsfig{file=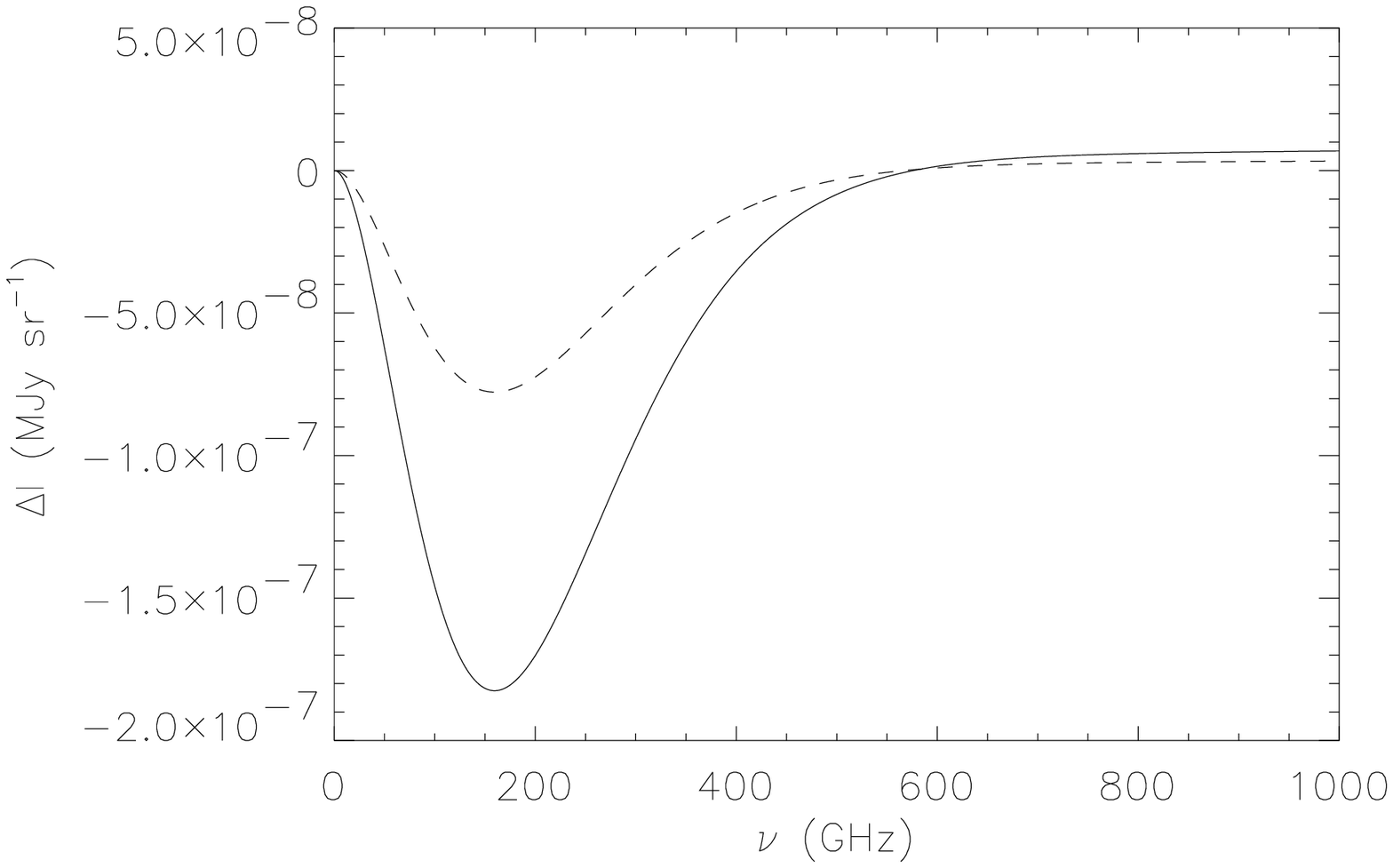,height=5.7cm,width=8.5cm,angle=0.0}
}
\end{center}
 \caption{Upper panel: non-thermal SZE in the Bullet Cluster produced in DM models in DME region (solid line) and in the DMW region (long dashed line), compared with WR models in MS and BS (dashed and dot-dashed lines), coincident with zero in this scale.
 Lower panel: non-thermal SZE in the Bullet Cluster produced in WR models in MS region (dashed line) and in the BS region (solid line).}
 \label{sz_4comp_nonterm}
\end{figure}

To compare the relative contributions of the secondary electron models in the different regions, we show in  Fig.\ref{sz_4comp_nonterm} the SZE produced in the DM and WR models.
The DM contribution results to be a factor $\sim 100$ smaller than the thermal one, while the WR contribution is smaller by a factor of $\sim10^6$.

In the light of these results, we conclude that the only important contributions come from the thermal components in the MS and in the BS, with the addition of the primary electron non-thermal effect found in the spectral analysis of the MS. This result is in agreement with the maps of Halverson et al. (2009) and Zemcov et al. (2010), where the SZE is located in the MS, with a small elongation towards the BS, and disagrees with the maps of Malu et al. (2010) and Plagge et al. (2010), where the SZE peak is shifted towards the DME region. We notice that a similar elongation seems to be present also in the radio maps (Shimwell et al. 2014), and that in this region there is a high temperature region with temperature higher than 20 keV (Markevitch et al. 2002). This last point can explain why the SZE appears to be shifted towards this region, being the SZE proportional to the Compton parameter $y\propto \int d\ell nkT$, while in this region the X-ray emission is low, so the density of the gas should be small (and it does not emerge also at larger X-ray energies, see Wik et al. 2014). Since in this region it is present the radio source A(L) that we already studied for its radio emission, it is also possible that this galaxy yields a contribution to the total SZE (see, e.g., Colafrancesco 2008, and Colafrancesco et al. 2013 for a study of the SZE in the radio galaxies lobes).

\begin{figure*}
\begin{center}
{
 \epsfig{file=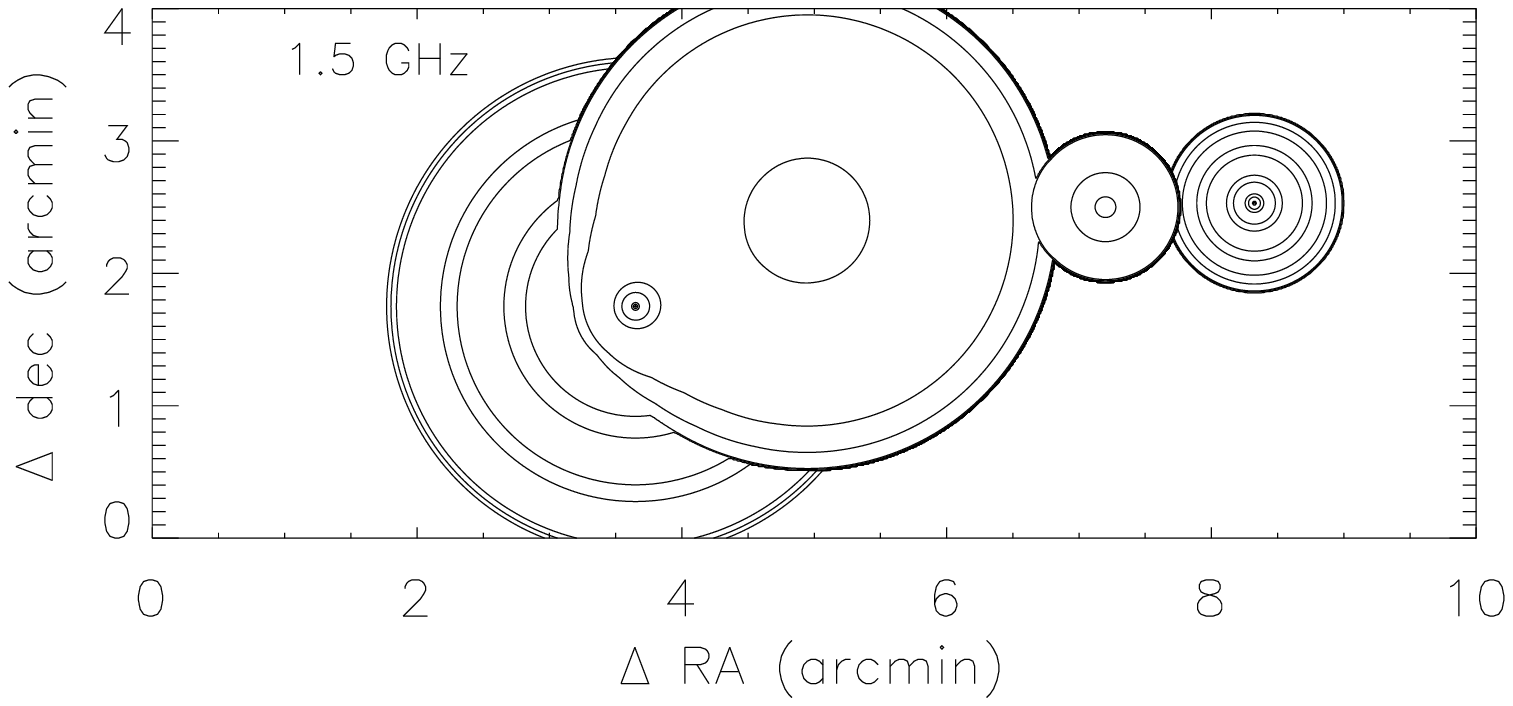,scale=0.5,angle=0.0}
 \epsfig{file=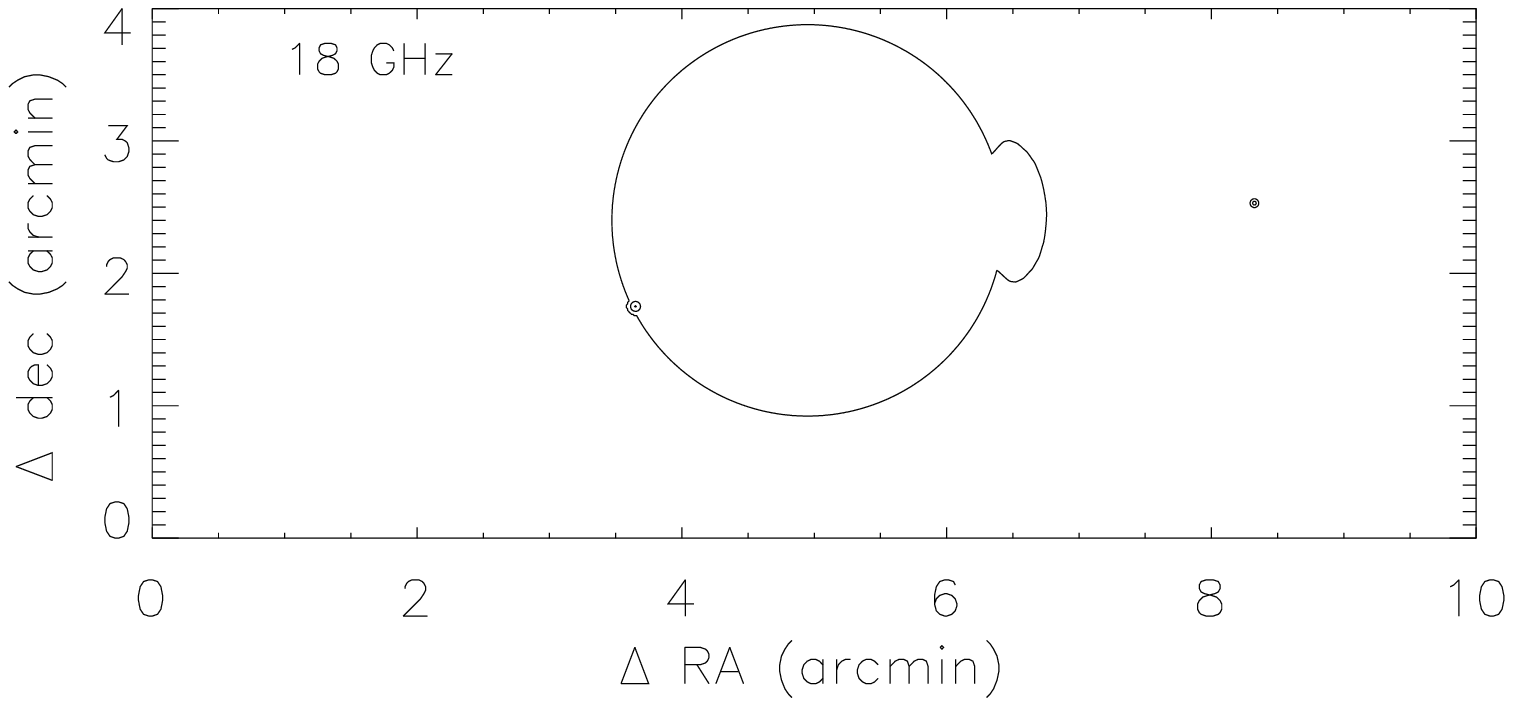,scale=0.5,angle=0.0}
 \epsfig{file=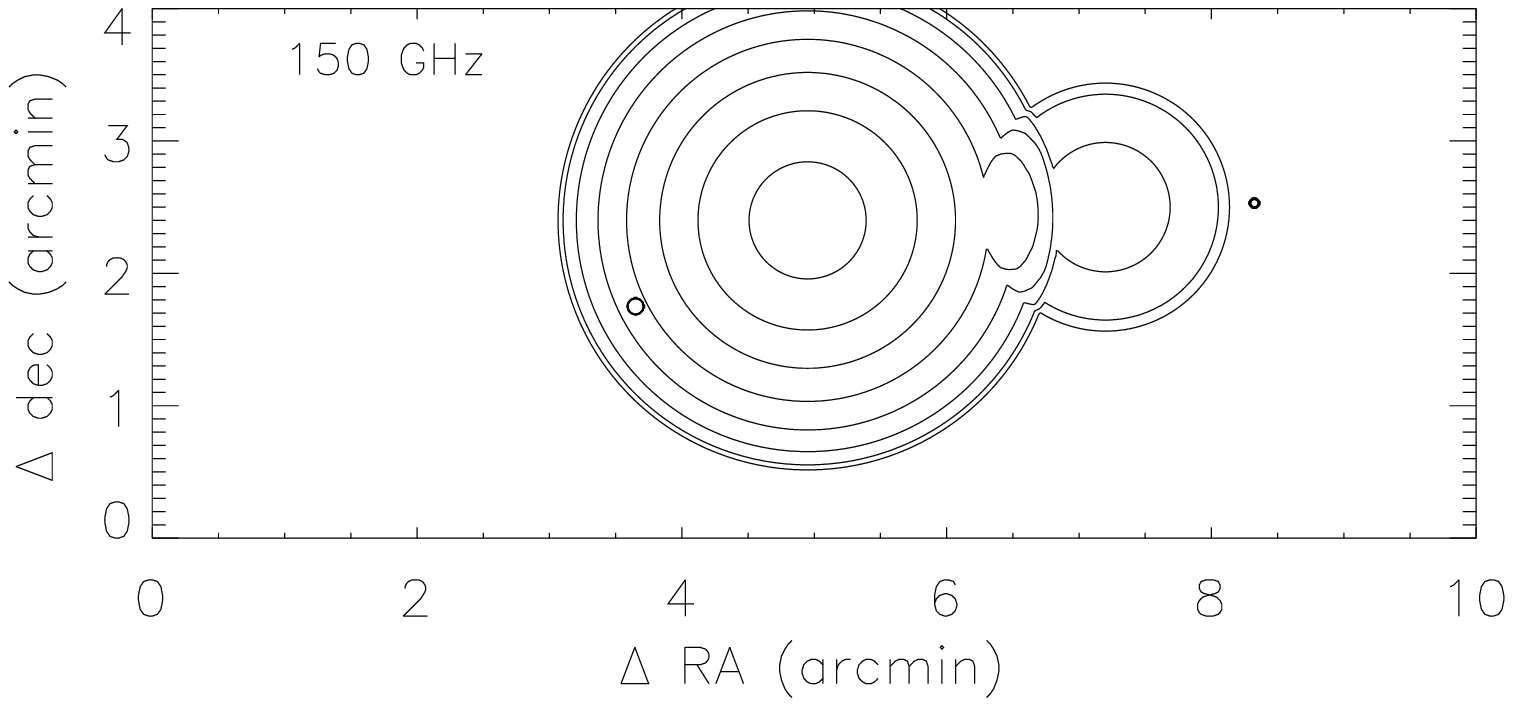,scale=0.5,angle=0.0}
 \epsfig{file=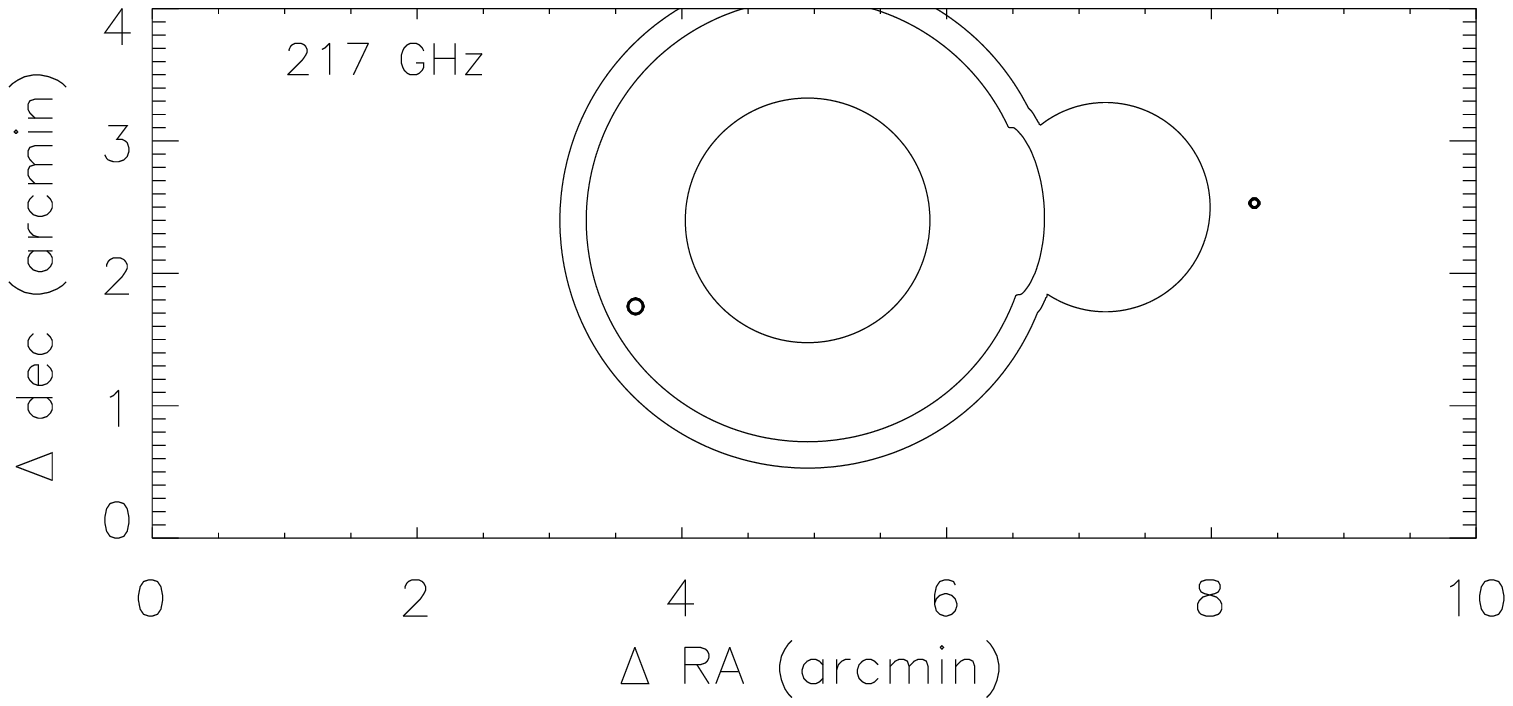,scale=0.5,angle=0.0}
  \epsfig{file=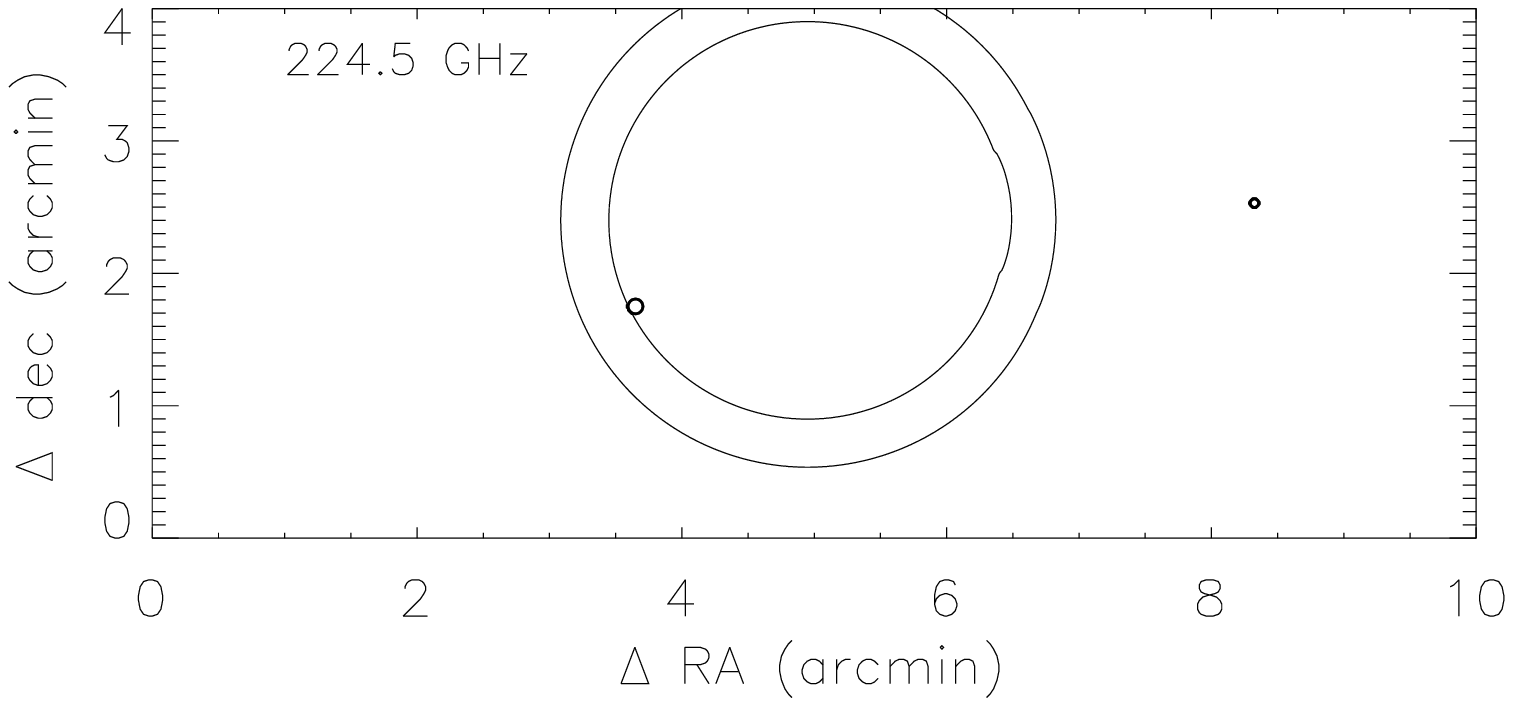,scale=0.5,angle=0.0}
 \epsfig{file=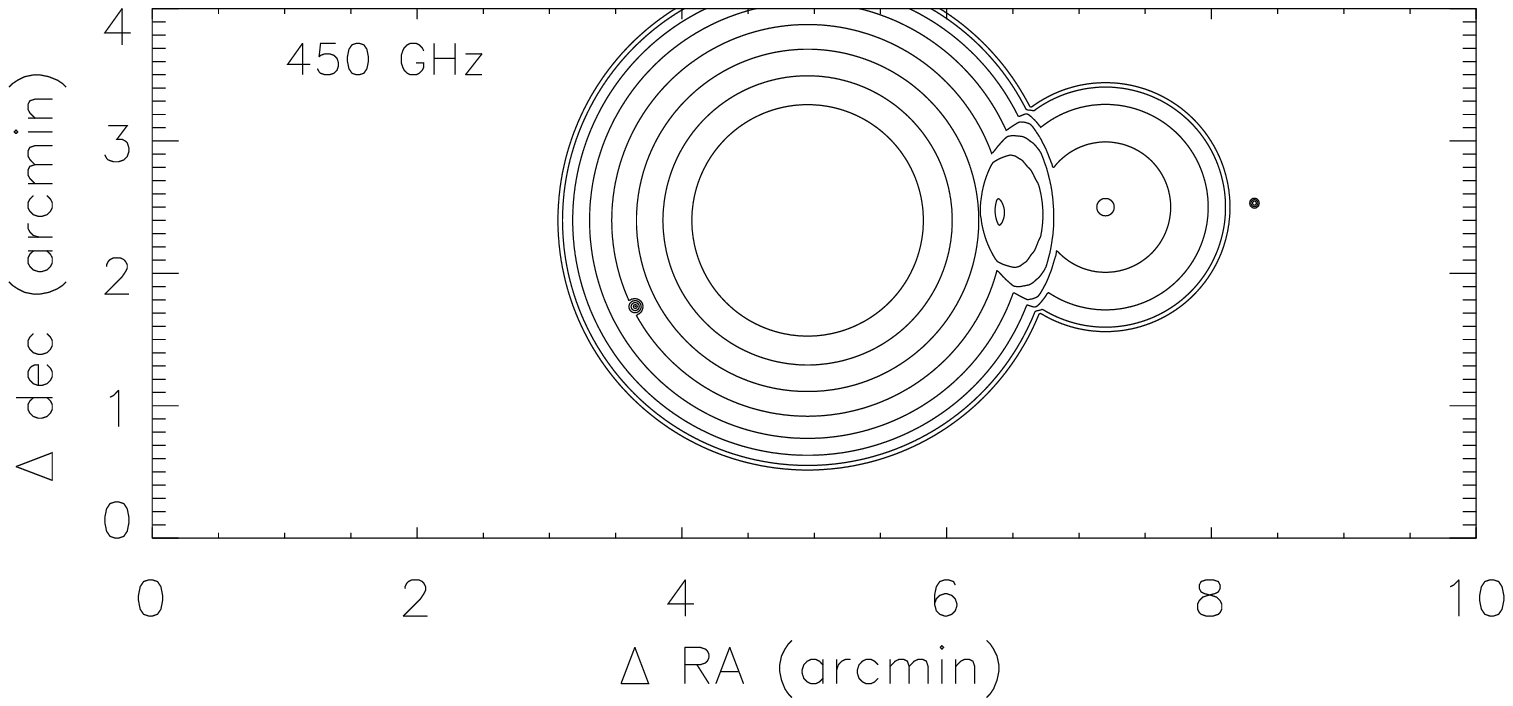,scale=0.5,angle=0.0}
  \epsfig{file=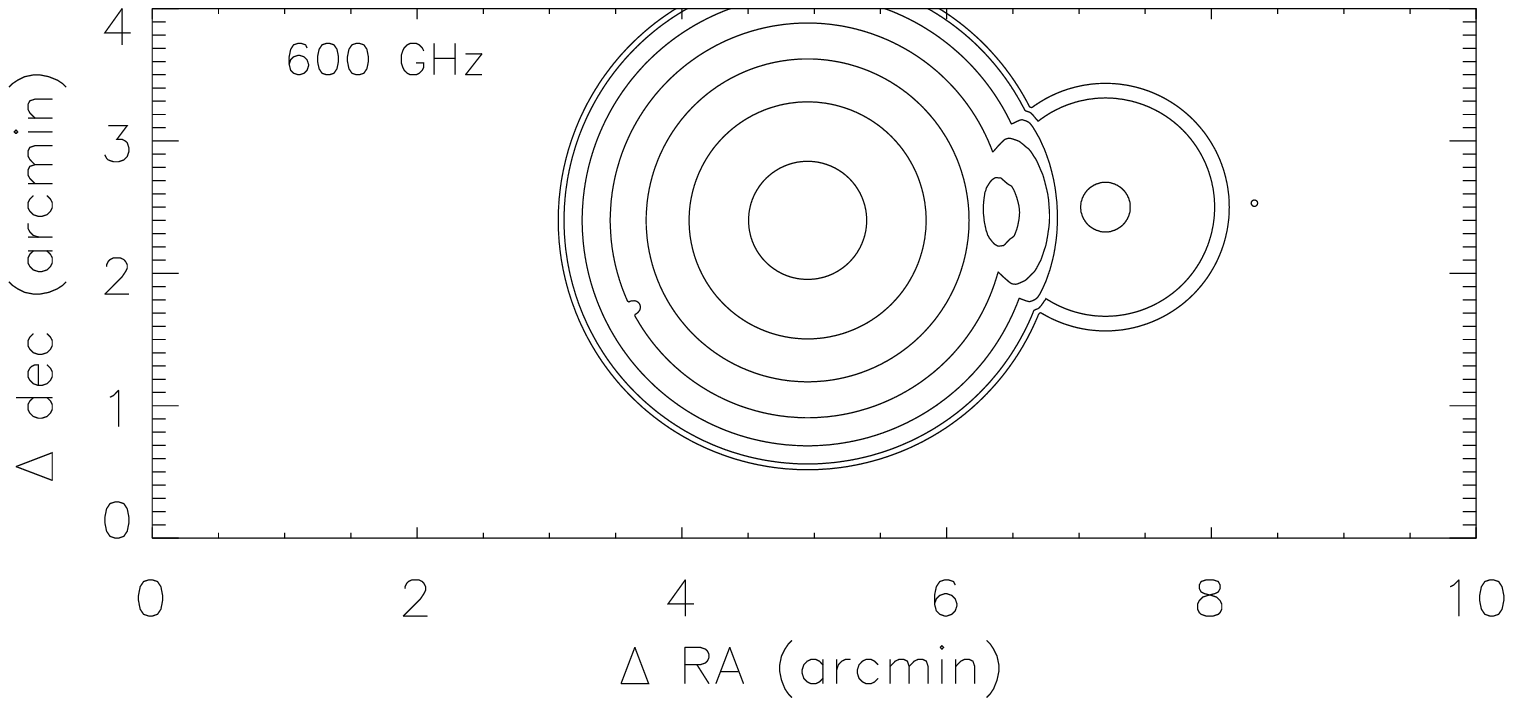,scale=0.5,angle=0.0}
 \epsfig{file=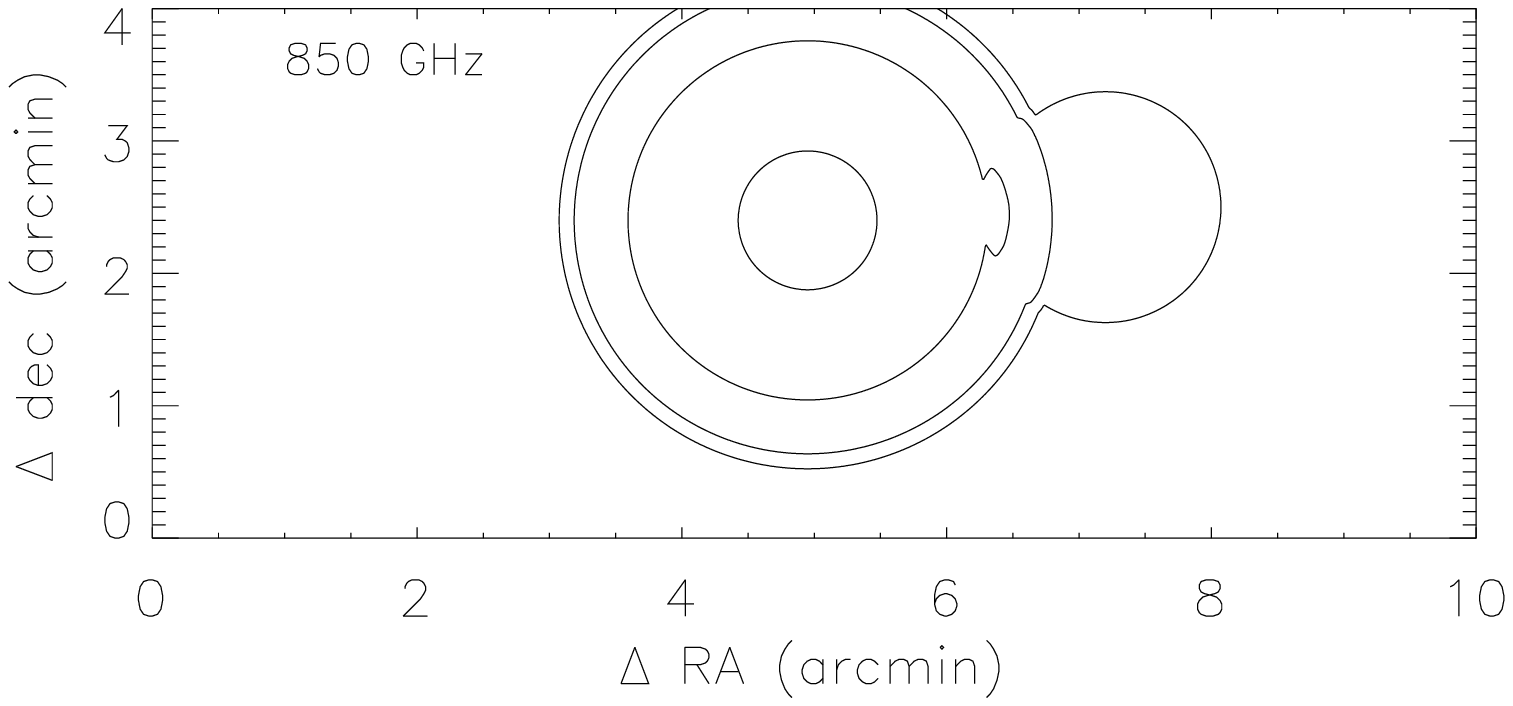,scale=0.5,angle=0.0}
}
\end{center}
 \caption{First panel: radio map at 1.5 GHz simulated with the surface brightness profiles plotted in the Fig.\ref{bril_4comp}. 
 Other panels: SZE maps at different frequencies simulated with the same models, and adding a non-thermal component in the MS produced by primary electrons as found in Sect.4.1. In all the panels, on the axes are the values of the coordinates differences (in arcmin) w.r.t. the origin, fixed in the point 06:58:50 -55:59:00 (J2000). In the first panel, contour levels correspond to: ($1\times10^{-6}$, $5\times10^{-6}$, $1\times10^{-5}$, $5\times10^{-5}$, $1\times10^{-4}$, $5\times10^{-4}$, $1\times10^{-3}$, $5\times10^{-3}$, $1\times10^{-2}$, $5\times10^{-2}$, $1\times10^{-1}$, $5\times10^{-1}$, 1, 5, 10, 50 and 100) Jy arcmin$^{-2}$. In the other panels, contour levels correspond (in absolute value) to: (1, 5, 10, 15, 20, 25, 30, 35) mJy arcmin$^{-2}$.}
 \label{mappe_bullet_varie}
\end{figure*}

In Fig.\ref{mappe_bullet_varie} we show some simulated maps of the SZE at several frequencies produced with the different models found in the previous sections, including for the MS the contribution from the non-thermal SZE from primary electrons. 

At 18 GHz (the frequency of the map of Malu et al. 2010) the SZE is very low, because this is a small frequency w.r.t. the frequency of the minimum; the SZE is negative and is dominated by the thermal effect in the MS, with a small elongation towards the BS and two point sources corresponding to the DM peaks.

At 150 GHz all the effects are negative (this frequency is near to the minimum of the SZE), and it is interesting to note that the combined effect of the SZE from MS and BS produces a secondary SZE peak in the middle of these two regions. This can be the origin of the elongated shape found in the map of Halverson et al. (2009) at the same frequency. The DM peaks appear to be very narrow, and they can be considered as point-like sources for most SZE experiments.
 
At 217 GHz, the frequency of the crossover point of the thermal effect in non-relativistic limit, the SZ effects are still negative and for the MS the dominant component is the non-thermal one for primary electrons, while there is still a residual thermal effect from the BS. We show also the case for 224.5 GHz, very close to the crossover frequency for the thermal effects if relativistic effects are considered, where the BS disappears almost entirely, and only the non-thermal component in the MS is present. This can be a good frequency where it is possible to test this model by detecting the non-thermal SZE (if we can distinguish it from a possible kinematic effect).

At 450 GHz the map is dominated by the thermal effects in the MS and BS, with the peak in the middle present also in this case. We note that, while the overall effect is positive, the DM peaks are negative, so they appear like two small holes in the map.

At 600 GHz (the frequency of the map of Zemcov et al. 2010) the SZE is positive, and it is dominated by the thermal peaks.

At 850 GHz the non-thermal effect in the MS is dominating w.r.t. the thermal effect, and the thermal effects in MS and BS appear as small corrections; the DM components are still negative and negligible w.r.t. the other effects.

We can compare these results with the expected performance of the Millimetron mission.
Millimetron is planned to have, at the wavelength of 300 $\mu$m and when operating in the single dish and medium spectral resolution mode, an angular resolution of $\approx 6$ arcsec and a sensitivity of $\approx 4$ $\mu$Jy for 3600 seconds of integration (Kardashev et al. 2015). In the SZE maps of Fig.\ref{mappe_bullet_varie} the lowest level contour is 1 mJy arcmin$^{-2}$; the flux integrated inside a circle with diameter of 6 arcsec for the lowest level contour of our maps is $\sim8$ $\mu$Jy, larger than the previous sensitivity limit. For this reason, we expect that Millimetron will be able to check our predictions by detecting the flux down to the lowest contour level we show in these maps, allowing hence to better constrain the properties of the different subhalos of the Bullet cluster. 

\section{Discussion and conclusions}

In this paper, we studied the complex structure of the radio halo in the Bullet Cluster by considering the properties of each of the baryonic and the DM subhalos observed with X-rays and gravitational lensing measurements. The radio emission from the MS is well explained by assuming a magnetic field of $\sim7.5$ $\mu$G, and a population of relativistic electrons centered on the MS. The radio observations do not allow to discriminate between a primary and a secondary origin for these electrons, and the forthcoming instruments in the X-ray and gamma-ray bands like Astro-H and CTA do not seem to be able to solve this problem, because the high distance of this cluster causes the emissions in these spectral bands to be smaller than the sensitivity limits of these instruments. Instead, SZE observations at high frequencies indicate that a primary origin is favoured.

The study of the other subhalos from radio data is more complicated, mainly because the available data do not separate the emissions coming from each of them, but consider the emission from a big region including all the subhalos and several point and extended sources. The radio spectrum of this region indicates that the presence of different emissions with different spectra is very probable. The combination of the emissions coming from the BS and the DME region can produce a spectrum similar to the observed one, but there are still various open issues:\\
\textit{i)} the curvature of the radio spectrum at $\nu>5$ GHz is difficult to be obtained for all the DM models we considered; this problem could be less important if we think that the procedure of sources removal in this region can be delicate;\\
\textit{ii)} the required annihilation cross section is very high ($\langle \sigma v \rangle \sim 8.1\times10^{-17}$ cm$^{3}$ s$^{-1}$), a factor of order of $\sim 10^8$ larger than the present upper limits, and probably it cannot be reproduced by a boosting factor due to DM substructures (see, e.g., Colafrancesco et al. 2015); anyway, if a higher (but reasonable) magnetic field of 5 $\mu$G is assumed, a similar spectrum can be obtained with a cross section larger by a factor of $10^3$  w.r.t. the present upper limits, and this difference can be recovered with a DM substructure boosting factor;\\
\textit{iii)} for these DM properties, an emission coming from the DMW region should also be observed, and this does not happen.

To better constrain the DM properties it would be important to perform better observations of each of the radio subhalos separately; this aim, both considering the sensitivity and the angular resolution properties, can be reached by forthcoming instruments like SKA, or its precursor MeerKAT.

Another possibility is that the radio emission coming from the DME region is really visible as a point source (due to the central peak of the DM distribution) that has been confused with the emission of the A(L) radio source, while its extended halo is included in the extended radio emission observed in this region. This hypothesis requires a smaller value of the cross section, and therefore also a smaller radio emission from the DMW region. To test this hypothesis, very detailed radio measurements, especially near the DME and DMW regions, are required.

A different possible explanation of the radio emission close to the DME region can be obtained by considering that near the DM peak there is also a big concentration of galaxies, so it is possible that the extended radio emission observed in this region is produced by the electrons emitted by the galaxies and diffusing in the ICM (Colafrancesco et al., in preparation). This is probably true at least for the emission coming from the regions in the north and the south of the BS, where the small extension of this subcluster does not predict a strong emission that instead is observed. We will test this hypothesis in a future paper.

Finally, the SZE appears to be the best tool to constrain the properties of the relativistic electrons in addition to the radio measurements. While the SZE produced by electrons produced by DM annihilation is probably too small to be observed, the study of the non-thermal SZE is anyway important in order to determine the properties of the electrons of baryonic origin, and to remove their contributions from the overall emission, favoring the study of the properties of the DM. Future instruments like Millimetron will be crucial to this aim.

\section*{Acknowledgments}

P.M. acknowledges support from the DST/NRF SKA post-graduate bursary initiative.
S.C. acknowledges support by the South African Research Chairs Initiative of the Department of Science and Technology and National Research Foundation and by the Square Kilometre Array (SKA). The authors thank the Reviewer and the Editors for useful comments and suggestions.

\bsp

\label{lastpage}

\end{document}